\documentclass[%
 longbibliography,
 reprint,
superscriptaddress,
 amsmath,amssymb,
 aps,
floatfix,
]{revtex4-2}

\usepackage{graphicx}
\usepackage{dcolumn}
\usepackage{bm}
\usepackage{color}
\usepackage{xr}
\usepackage[normalem]{ulem}
\usepackage[dvipsnames]{xcolor}

\begin{document}

\title{Modeling non-Poissonian temporal hypergraphs by Markovian node dynamics}

\author{Hang-Hyun Jo}
\email{h2jo@catholic.ac.kr}
\affiliation{Department of Physics, The Catholic University of Korea, Bucheon, Republic of Korea}

\author{Naoki Masuda}
\email{naokimas@umich.edu}
\affiliation{Gilbert S. Omenn Department of Computational Medicine and Bioinformatics, University of Michigan, Ann Arbor, MI, USA}
\affiliation{Department of Mathematics, University of Michigan, Ann Arbor, MI, USA}
\affiliation{Center for Computational Social Science, Kobe University, Kobe, Japan}

\date{\today}

\begin{abstract}
Temporal hypergraphs capture time-resolved group interactions among nodes. Empirical data support that time-stamped group interactions show bursty event sequences and non-trivial temporal correlations. In the present study, we introduce node-driven temporal hypergraph models in which each node stochastically alternates between low- and high-activity states, and a hyperedge produces time-stamped events with a probability that depends on the number of high-state nodes in the hyperedge. For two event-generation rules, we analytically derive interevent time distributions and autocorrelation functions of event sequences, both for hyperedges and nodes. Despite Markovian node-state dynamics, the induced event processes become mixtures of Poissonian, short-tailed components, resulting in longer-tailed interevent time distributions and slowly decaying autocorrelation. The theory further shows the dependence of these features on the size of hyperedge, which largely agrees with various empirical data. We expect our models to provide a simple, interpretable framework for connecting individual-level activity fluctuations to the timing patterns observed in real group interactions.
\end{abstract}

\maketitle


\section*{Introduction}

In many real-world networked systems, edges connecting pairs of units are traditionally assumed to be static but in fact vary over time. Those edges may appear and disappear, or may be active only intermittently over time. Temporal network analysis provides a powerful theoretical framework and various algorithms for such data~\cite{Holme2012Temporal, Holme2015Modern, Holme2019Temporal, Masuda2020Guide}. Temporal network analysis has extensively modified our understanding of the structure and dynamics of networks that had been traditionally studied for static networks. The structure of temporal networks has been characterized by, e.g., temporal motifs~\cite{Kovanen2011Temporal, Paranjape2017Motifs, Liu2023Temporal}, temporal community structure~\cite{Fortunato2010Community, Bazzi2016Community}, non-Poissonian bursty temporal patterns on edges and nodes~\cite{Barabasi2005Origin, Vazquez2006Modeling, Karsai2018Bursty}, temporal correlation of edges~\cite{Goh2008Burstiness, Saramaki2015Seconds}, and concurrency and memory shapes across different edges~\cite{Masuda2021Concurrency, Williams2022Shape}. Collective dynamics taking place on temporal networks has also been extensively analyzed; examples include diffusion~\cite{Perra2012Random, Starnini2012Random, Masuda2013Temporal}, synchronization~\cite{Kohar2014Synchronization, Zhang2021Designing}, epidemic spreading~\cite{Karsai2011Small, VanMieghem2013NonMarkovian, Rosvall2014Memory, Masuda2017Introduction}, and network control~\cite{Li2017Fundamental}.

Many real-world systems, from biological to social systems, are characterized by the presence of group interactions, which can not be decomposed into a linear combination of dyadic ties. Hypergraphs, rather than conventional networks that are limited to pairwise interactions, are capable of describing systems and data with such higher-order interactions; a hyperedge represents an interaction among an arbitrary number of nodes~\cite{Lambiotte2019Networks, Battiston2020Networks, Battiston2021Physics, Bianconi2021Higherorder, Majhi2022Dynamics, Bick2023What, FerrazDeArruda2024Contagion, Battiston2026Collective}. Recently, analyses of empirical hypergraphs have revealed that higher-order interactions are typically bursty~\cite{Cencetti2021Temporal, Ceria2023Temporaltopological, Iacopini2024Temporal, Mancastroppa2025Emerging} (as pairwise interaction are \cite{Barabasi2005Origin, Vazquez2006Modeling, Karsai2018Bursty}), that they display non-trivial temporal correlations~\cite{Gallo2024Higherorder, Iacopini2024Temporal, BartPeters2025Higherorder}, and that temporal and topological patterns of higher-order events are also correlated~\cite{Ceria2023Temporaltopological}. In this sense, many empirical hypergraphs are temporal hypergraphs. Furthermore, non-trivial temporal patterns in higher-order interactions affect dynamics on hypergraphs such as epidemic processes~\cite{Chowdhary2021Simplicial, St-Onge2021Universal}, consensus dynamics~\cite{Neuhauser2021Consensus}, and the evolution of cooperation~\cite{Wang2026Strategy}, as they do for dynamics on networks, i.e., in the case of dyadic interactions. 
Nonetheless, we acknowledge recent studies showing that dynamical systems on hypergraphs can be transformed into those on conventional networks without loss of information in many cases~\cite{Peixoto2026Graphs, Lacasa2026Equivalence}.

We focus on modeling the structure of temporal hypergraphs, especially burstiness of event sequences in hypergraphs. A variety of models have already been introduced to describe dynamics of group interactions. In the higher-order activity-driven (HAD) models~\cite{Petri2018Simplicial, DiGaetano2024Percolation, Han2024Probabilistic}, at every discrete time, group interactions among nodes are randomly generated as a function of the predetermined activity of nodes. While the HAD models are advantageous in analytical tractability, they do not account for bursty, non-Poissonian activity patterns present in empirical time-dependent group-interaction data, at least in its original form. More involved computational models of temporal hypergraphs, including variants of HAD models, have been proposed for producing memory within and across group sizes~\cite{Gallo2024Higherorder}, patterns of group transitions~\cite{Iacopini2024Temporal}, individual node dynamics~\cite{Mancastroppa2025Emerging}, and the emergence of hypercores~\cite{Mancastroppa2024Structural}. Other models of temporal hypergraphs allow statistical inference of the model parameters~\cite{Lerner2025Modeling}. However, temporal properties of these computational models are difficult to examine in analytical terms.

In this study, we propose a family of temporal hypergraph models whose non-Poissonian temporal properties can be analytically studied. Although there are various analytically tractable modeling for temporal networks~\cite{Masuda2020Guide, Perra2012Activity, Vestergaard2014How, Zhang2017Random, Hiraoka2020Modeling} and hypergraphs~\cite{Petri2018Simplicial, DiGaetano2024Percolation, Han2024Probabilistic}, our models are node-activity models that have been used for modeling both static~\cite{Goh2001Universal, Caldarelli2002ScaleFree, Chung2002Average, Boguna2003Class, Masuda2004Analysis} and temporal~\cite{Perra2012Activity, Petri2018Simplicial, DiGaetano2024Percolation, Han2024Probabilistic, Mancastroppa2025Emerging} networks. In our models, we assume that nodes' activities vary over time in a Markovian manner, inspired by previous models for temporal networks~\cite{Malmgren2008Poissonian, Malmgren2009Universality, Karsai2012Universal, FonsecaDosReis2020Generative, Hartle2025Autocorrelation}, and that a hyperedge is activated if many nodes belonging to it are simultaneously activated. We derive the distribution of interevent times (IETs) and the autocorrelation function (ACF) of time-stamped events for our models, and compare results with statistics of empirical data.

\section*{Results}

\subsection*{Model}\label{sec:model}

We consider a substrate static hypergraph with $N$ nodes and $E$ hyperedges, on which we generate time-stamped events. We denote the $i$th node (with $i \in \{1, \ldots, N\}$) by $v_i$ and the $j$th hyperedge (with $j \in \{1, \ldots, E \}$) by $e_j$.

\begin{figure*}[!t]
    \centering
    \includegraphics[width=\textwidth]{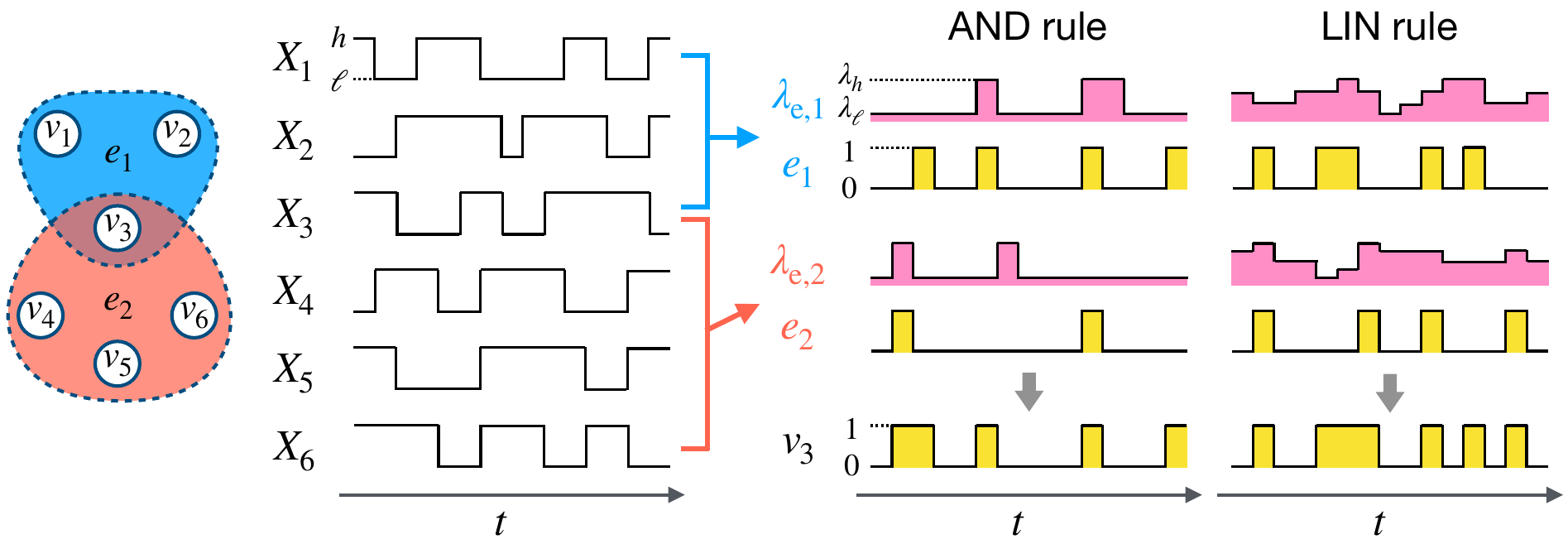}
    \caption{Schematic of the dynamics of six nodes, $v_1,\ldots,v_6$, and two hyperedges, $e_1$ and $e_2$. The two hyperedges share one node, $v_3$. Variable $X_i(t)$ for $i \in \{ 1,\ldots,6 \}$ denotes the state of $v_i$ at time $t$, which is either $h$ or $\ell$. Variable $\lambda_{{\rm e},j}(t)$ for $j \in \{1,2 \}$ represents the event probability on the $j$th hyperedge at time $t$ [Eqs.~\eqref{eq:event_rate_and} and~\eqref{eq:event_rate_lin} for AND and LIN rules, respectively]. Note that $0<\lambda_{\ell}\leq \lambda_{{\rm e},j}\leq \lambda_h$. Variables $e_1(t)$, $e_2(t)$, and $v_3(t)$ represent the time series of events on two hyperedges and node $v_3$; at any $t$, they are equal to $1$ when there is an event and $0$ when there is no event. Equation~\eqref{eq:vt_ejt_relation} determines $v_3(t)$ from $e_1(t)$ and $e_2(t)$.}
    \label{fig:schematic}
\end{figure*}

We assume that the state of each $i$th node, denoted by $X_i\in\{h, \ell\}$, flips between $h$ (high) and $\ell$ (low) over time (see Fig.~\ref{fig:schematic}) . The dynamics of $X_i$ is independent and statistically identical across all nodes and modeled by the same stochastic process. We also assume that the state of the hyperedge at any time is determined by the states of the nodes belonging to the hyperedge. This assumption is an extension of previous models of temporal networks \cite{FonsecaDosReis2020Generative, Hartle2025Autocorrelation}. Each hyperedge produces an event with a probability that is a function of the hyperedge's state.

Although we can define the rest of our models in both continuous and discrete time, we describe them in discrete time. See Supplementary Note~1 for the derivation of IET distributions in continuous time for completeness.

We consider two rules, called the AND and linear (LIN), to determine the state and event probability of the hyperedges. Under the AND rule~\cite{FonsecaDosReis2020Generative, Hartle2025Autocorrelation}, the hyperedge is in the $h$ state if and only if all nodes belonging to the hyperedge are in the $h$ state. Otherwise, it is in the $\ell$ state. When the hyperedge is in the $h$ or $\ell$ state, an event occurs on the hyperedge with a probability $\lambda_h$ or $\lambda_{\ell}$ ($<\lambda_h$), respectively. In other words, an event occurs on the hyperedge with probability
\begin{align}
    \lambda_{\rm e}(m,m_h) \equiv \delta_{m,m_h}\lambda_h +(1-\delta_{m,m_h})\lambda_{\ell},
    \label{eq:event_rate_and}
\end{align}
where $m$ is the size of the hyperedge, $m_h \in \{0,\ldots,m\}$ is the number of nodes in the $h$ state in the hyperedge, and $\delta$ is the Kronecker delta. 

Under the LIN rule, the event probability on the hyperedge is given by
\begin{align}
    \lambda_{\rm e}(m,m_h) \equiv \frac{m_h}{m}\lambda_h+\left(1-\frac{m_h}{m}\right)\lambda_{\ell}.
    \label{eq:event_rate_lin}
\end{align}
Note that the LIN rule is similar to the IND model in Ref.~\cite{FonsecaDosReis2020Generative} in the sense that the event probability on the (hyper)edge is a linear function of $m_h$. Both AND and LIN rules reduce to the Bernoulli process (i.e., discrete variant of the Poisson process) when $\lambda_h=\lambda_{\ell}$.

At any time, after an event may have occurred on each hyperedge $e$ (with probability $\lambda_{\rm e}$), each node in the $h$ state is assumed to flip to the $\ell$ state with probability $r_{h \ell}$ ($>0$). Each node in the $\ell$ state flips to the $h$ state with probability $r_{\ell h}$ ($>0$).

\subsection*{Interevent time distributions}\label{sec:IET}

Our goal of this section is to theoretically derive the interevent time (IET) distribution for nodes and hyperedges. The IET is defined as the time difference between two consecutive time-stamped events occurring on a node or hyperedge.

We first describe the dynamics of a single node's state. Let us denote the probability that a node is in the $h$ state at time step $t$ by $p_h(t)$. A node is in the $\ell$ state at time step $t$ with probability $1-p_h(t)$. The recursive equation for $p_h(t)$ and its solution are given by~\cite{Hartle2025Autocorrelation}
\begin{align}
    p_h(t)&=p_h(t-1)(1-r_{h \ell})+[1-p_h(t-1)]r_{\ell h}\notag\\
    &=p_h^* + [p_h(0)-p_h^*]\eta^t,
    \label{eq:ph_t}
\end{align}
where
\begin{align}
    p_h^*\equiv\frac{r_{\ell h}}{r_{\ell h}+r_{h \ell}}
    \label{eq:ph_rr}
\end{align}
is the probability that a node is in the $h$ state at equilibrium, and
\begin{equation}
    \eta\equiv 1-r_{\ell h}-r_{h \ell}.
    \label{eq:eta-def}
\end{equation}
The node is in the $\ell$ state with probability $1-p_h^*$ at equilibrium. For notational simplicity, we write $p_h$ for $p_h^*$ in the following text. 

\subsubsection*{IET distribution for hyperedges}\label{subsec:IET_hyper}

Here we derive the analytical solution of the IET distribution for the hyperedge of size $m$. We describe the hyperedge's activity by $e(t)$, which is defined to be equal to $1$ when an event occurs at time $t$; otherwise, $e(t) = 0$ (see Fig.~\ref{fig:schematic}). Whether or not an event occurs at time $t$ depends on the number of nodes belonging to the hyperedge that are in the $h$ state at time $t$, which we denote by $m_h(t)$. The IET distribution for the hyperedge can be expressed as
\begin{align}
    P_{\rm e}(\tau)=&\Pr[e(\tau)=1, e(\tau-1)=0,\ldots, e(1)=0|e(0)=1] \notag\\
    =&\sum_{n_0,\ldots,n_{\tau}=0}^m \Pr[e(\tau)=1|m_h(\tau)=n_{\tau}] \notag\\ &\times \left[\prod_{t=1}^{\tau-1} \Pr[m_h(t+1)=n_{t+1}|m_h(t)=n_t] \right.\notag\\
    &\times\Pr[e(t)=0|m_h(t)=n_t] \bigg]\notag \\ 
    &\times \Pr[m_h(1)=n_1|m_h(0)=n_0] \notag\\
    &\times \Pr[m_h(0)=n_0|e(0)=1],
    \label{eq:Ptau_e_discrete}
\end{align}
where $\Pr[\cdot | \cdot]$ represents the conditional probability. For any $t$, we obtain
\begin{align}
    \Pr[e(t)=1|m_h(t)=n_t] &= \lambda_{\rm e}(m,n_t), \label{eq:Pr_1n}\\
    \Pr[e(t)=0|m_h(t)=n_t] &= 1-\lambda_{\rm e}(m,n_t)\equiv \bar\lambda_{\rm e}(m,n_t),
    \label{eq:Pr_0n}
\end{align}
where $\lambda_{\rm e}$ is given by Eqs.~\eqref{eq:event_rate_and} and~\eqref{eq:event_rate_lin} for the AND and LIN rules, respectively. 

To compute $\Pr[m_h(t+1)=n_{t+1}|m_h(t)=n_t]$ in Eq.~\eqref{eq:Ptau_e_discrete}, we write the master equation of the random walk that $m_h(t)$ obeys. Specifically, the probability that there are $n$ nodes in the $h$ state at time $t$, denoted by $Q_n(t)$, obeys
\begin{align}
    Q_n(t+1)=\sum_{n'=0}^m W_{{\rm e},nn'}Q_{n'}(t),
    \label{eq:Pnt_master}
\end{align}
where $W_{\rm e}=(W_{{\rm e},nn'})$ is the $(m+1)\times(m+1)$ transition probability matrix from $n'$ to $n$. Its entries are given by
\begin{align}
    W_{{\rm e},nn'}=&\sum_{u=0}^{n'}\sum_{u'=0}^{m-n'} {n' \choose u}r_{h\ell}^u \bar r_{h\ell}^{n'-u} {m-n' \choose  u'}r_{\ell h}^{u'} \bar r_{\ell h}^{m-n'-u'}\notag\\
    &\times \delta_{n,n'-u+u'},
    \label{eq:Wenn'_define}
\end{align}
where $n' \choose u$ is the binomial coefficient, $\bar r_{h \ell}\equiv 1-r_{h \ell}$, and $\bar r_{\ell h}\equiv 1-r_{\ell h}$. Each summand on the right-hand side of Eq.~\eqref{eq:Wenn'_define} is a product of three quantities. First, ${n' \choose u}r_{h\ell}^u \bar r_{h\ell}^{n'-u}$ is the probability of choosing $u$ nodes that change their states from $h$ to $\ell$, among the $n'$ nodes in the $h$ state. Second, ${m-n' \choose  u'}r_{\ell h}^{u'} \bar r_{\ell h}^{m-n'-u'}$ is the probability of choosing $u'$ nodes that change their states from $\ell$ to $h$, among the $m-n'$ nodes in the $\ell$ state. Third, $\delta_{n,n'-u+u'}$ imposes that the resultant number of nodes in the $h$ state, i.e., $n'-u+u'$, must be $n$. Then, one obtains
\begin{align}
    \Pr[m_h(t+1)=n_{t+1}|m_h(t)=n_t]=W_{{\rm e},n_{t+1}n_t}.
    \label{eq:Pr_nt+1nt}    
\end{align}

Finally, to compute $\Pr[m_h(0)=n_0|e(0)=1]$ in Eq.~\eqref{eq:Ptau_e_discrete}, we use the Bayes' theorem~\cite{Ross2014First} to obtain
\begin{align}
    &\Pr[m_h(0)=n_0|e(0)=1]\notag\\
    &=\frac{\Pr[e(0)=1|m_h(0)=n_0] \Pr[m_h(0)=n_0]}{\Pr[e(0)=1]}\notag\\
    &=\frac{\lambda_{\rm e}(m,n_0) f_{m n_0}}{\Omega_{\rm e}}.
    \label{eq:Pr_n1}
\end{align}
We have assumed the equilibrium for $n_0$ at $t=0$ and used $\Pr[m_h(0)=n_0] = f_{mn_0}$, where we define for $0\leq n'\leq n$
\begin{align}
    f_{nn'}\equiv {n\choose n'}p_h^{n'}\bar p_h^{n-n'},
    \label{eq:fmn}
\end{align}
and $\bar p_h\equiv 1-p_h$. We have also defined $\Omega_{\rm e}\equiv \Pr[e(0)=1]$, which is the average event probability on the hyperedge. Specifically, we have
\begin{align}
    \Omega_{\rm e}=\sum_{n_0=0}^m f_{m n_0}\lambda_{\rm e}(m,n_0).
    \label{eq:Omega_e}
\end{align}

In sum, once the hyperedge's size $m$ and the rule for generating events on it, i.e., AND or LIN, are given, by combining Eqs.~\eqref{eq:Pr_1n},~\eqref{eq:Pr_0n},~\eqref{eq:Wenn'_define},~\eqref{eq:Pr_nt+1nt}, and~\eqref{eq:Pr_n1}, one can derive the exact solution of $P_{\rm e}(\tau)$ in Eq.~\eqref{eq:Ptau_e_discrete} as a function of $r_{\ell h}$, $r_{h\ell}$, $\lambda_{\ell}$, and $\lambda_h$ for each $m$. To proceed further, we assume that $r_{\ell h},r_{h \ell}\ll 1$, i.e., that the transitions of nodes between the $h$ and $\ell$ states are rare. As we show in Supplementary Note~2, one then obtains an approximate analytical solution of the IET distribution as
\begin{align}
    P_{\rm e}(\tau) \approx \frac{1}{\Omega_{\rm e}}\sum_{n=0}^m f_{mn}\lambda_{\rm e}(m,n)^2 \bar\lambda_{\rm e}(m,n)^{\tau-1}.
    \label{eq:Ptau_e_approx}
\end{align}
Equation~\eqref{eq:Ptau_e_approx} is a weighted sum of geometric distributions with different probabilities ranging from $\lambda_{\rm e}(m,0)$ to $\lambda_{\rm e}(m,m)$. For both AND and LIN rules, $\lambda_{\rm e}(m,0)=\lambda_{\ell}$ and $\lambda_{\rm e}(m,m)=\lambda_h$. Because $\lambda_{\ell}<\Omega_{\rm e}<\lambda_h$ from Eq.~\eqref{eq:Omega_e}, our model has a heavier tail in the IET distribution than the corresponding Bernoulli process (i.e., discrete-time counterpart of Poisson processes) with the same mean event probability $\Omega_{\rm e}$~\cite{Yannaros1994Weibull, Feldmann1998Fitting, Masuda2022Gillespie}, whose IET distribution is a geometric distribution given by
\begin{align}
    P_{\rm e}^{\rm B}(\tau)=\Omega_{\rm e} (1-\Omega_{\rm e})^{\tau-1}.
    \label{eq:Ptau_e_avg}
\end{align}

\begin{figure}[!t]
    \centering
    \includegraphics[width=\columnwidth]{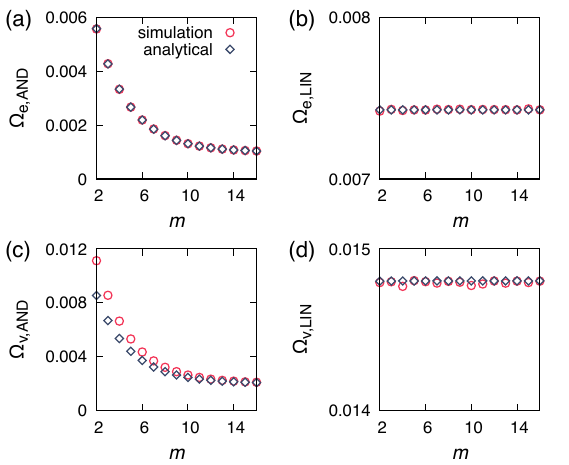}
    \caption{Average event probabilities $\Omega$ as a function of hyperedge size $m$. (a) Hyperedge, AND rule. (b) Hyperedge, LIN rule. (c) Node, AND rule. (d) Node, LIN rule. We have computed numerical results (red circles) with $r_{\ell h}=5\times 10^{-5}$, $r_{h \ell}=2\times 10^{-5}$, $\lambda_h=10^{-2}$, and $\lambda_{\ell}=10^{-3}$, following the numerical procedures described in the Methods section. For the analytical results (dark blue diamonds), we numerically calculate Eqs.~\eqref{eq:Omega_e_and_m},~\eqref{eq:Omega_e_lin_m},~(S41), and~(S45).
    }
    \label{fig:omega}
\end{figure}

Under the AND rule, we use Eq.~\eqref{eq:event_rate_and} to obtain
\begin{align}
    P_{\rm e,AND}(\tau)\approx \frac{1}{\Omega_{\rm e,AND}}\left[
    p_h^m\lambda_h^2 \bar\lambda_h^{\tau-1}
    +(1-p_h^m)\lambda_{\ell}^2 \bar\lambda_{\ell}^{\tau-1}
    \right],
    \label{eq:Ptau_e_and_m}
\end{align}
where
\begin{align}
    \Omega_{\rm e,AND}=p_h^m \lambda_h +(1-p_h^m)\lambda_{\ell},
    \label{eq:Omega_e_and_m}
\end{align}
$\bar\lambda_h\equiv 1-\lambda_h$, and $\bar\lambda_{\ell}\equiv 1-\lambda_{\ell}$. We remark that $\Omega_{\rm e,AND}$ decreases with $m$ [see Fig.~\ref{fig:omega}(a)]. This tendency is because a large $m$ reduces $p_h^m$, i.e., the probability of the hyperedge being in the $h$ state, implying that most events occur with probability $\lambda_{\ell}$ when $m$ is large. Consistent with this explanation, the event-generation process is reduced to the Bernoulli process as $m\to\infty$, resulting in the geometric IET distribution with event probability $\lambda_{\ell}$.

Under the LIN rule, we use Eq.~\eqref{eq:event_rate_lin} to obtain
\begin{align}
    P_{\rm e,LIN}(\tau)\approx & \frac{1}{\Omega_{\rm e,LIN}}
    \sum_{n=0}^m f_{mn}\left[\frac{n}{m}\lambda_h+\left(1-\frac{n}{m}\right)\lambda_{\ell}\right]^2 \notag\\
    & \times \left[\frac{n}{m}\bar\lambda_h+\left(1-\frac{n}{m}\right)\bar\lambda_{\ell}\right]^{\tau-1},
    \label{eq:Ptau_e_lin_m}
\end{align}
where
\begin{align}
    \Omega_{\rm e,LIN}=p_h\lambda_h+\bar p_h\lambda_{\ell}.
    \label{eq:Omega_e_lin_m}
\end{align}
We remark that $\Omega_{\rm e,LIN}$ in Eq.~\eqref{eq:Omega_e_lin_m} is independent of $m$ [see Fig.~\ref{fig:omega}(b)], whereas the shape of $P_{\rm e,LIN}(\tau)$ in Eq.~\eqref{eq:Ptau_e_lin_m} depends on $m$. In the limit of $m\to\infty$, the most probable value of $m_h$, i.e., the number of nodes in the $h$ state, approaches $m/2$, implying that most events occur with probability $(\lambda_h+\lambda_{\ell})/2$. Similar to the case of the AND rule, the event-generation process becomes the Bernoulli process as $m\to\infty$, leading to the geometric IET distribution with probability $(\lambda_h+\lambda_{\ell})/2$.

We define the duration of activity, denoted by $d$, as the number of consecutive events (i.e., $e(t)=1$) separated by non-events (i.e., $e(t)=0$). We have obtained the approximate duration distribution $P_{\rm e}(d)$ under both AND and LIN rules as in Eqs.~(S28) and~(S29). See Supplementary Note~3 for details.

\subsubsection*{IET distribution for nodes}\label{subsec:IET_node}

Now we derive the IET distribution for a node belonging to $k$ hyperedges of sizes $\vec m=(m_1,\ldots,m_k)$. Let $v(t)=1$ if at least one of the $k$ hyperedges generates an event at time $t$. Otherwise, we set $v(t)=0$ (see Fig.~\ref{fig:schematic}). We further denote the presence (i.e., $=1$) or absence (i.e., $=0$) of an event at time $t$ on the $j$th hyperedge (with $j\in\{1,\ldots,k\}$) by $e_j(t)$. Then, one obtains
\begin{align}
    v(t)=1-\prod_{j=1}^k [1-e_j(t)].
    \label{eq:vt_ejt_relation}
\end{align}
The behavior of $e_j(t)$ has been fully described in the previous section; the value of $e_j(t)$ is determined by the number of $h$-state nodes among $m_j$ nodes, which we denote by $m_{jh}$. Specifically, the probability that an event occurs on the $j$th hyperedge, $\lambda_{{\rm e},j}(m_j,m_{jh})$, is a function of $m_{jh}$.

To study the behavior of $v(t)$, we need to describe the dynamics of the states of the focal node and its neighbors altogether. To do this, we consider the focal node as an additional ``hyperedge" of its own with index $0$. We denote the size of this $0$th hyperedge by $\mu_0$, and it is equal to $1$. The number of nodes in the $j$th hyperedge (with $j\in\{1,\ldots,k\}$) excluding the focal node is then $m_j-1$. For notational convenience, we write $\mu_j \equiv m_j - 1$ and $\vec\mu \equiv (\mu_0,\mu_1,\ldots,\mu_k)$. For example, node $v_3$ in Fig.~\ref{fig:schematic} belongs to two hyperedges of sizes $\vec m =(3,4)$, yielding $\vec\mu=(1,2,3)$. Next, we consider the number of $h$-state nodes in each hyperedge, including the aforementioned $0$th hyperedge; these numbers are collectively denoted by a vector $\vec\mu_h\equiv (\mu_{0h},\mu_{1h},\ldots, \mu_{kh})$. Note that $0\leq \mu_{jh}\leq \mu_j$ for $j\in\{0,1,\ldots,k\}$ and that $m_{jh} = \mu_{0h}+\mu_{jh}$ for $j\in\{1,\ldots,k\}$. Then, using Eq.~\eqref{eq:vt_ejt_relation}, we obtain the event probability on the node in terms of event probabilities on the hyperedges as
\begin{align}
    \lambda_{\rm v}(\vec \mu,\vec \mu_h) = 1-\prod_{j=1}^k \bar\lambda_{{\rm e},j}(1+\mu_j,\mu_{0h}+\mu_{jh}).
    \label{eq:event_node_edge}
\end{align}
For both AND and LIN rules,  $\lambda_{\ell}\leq \lambda_{{\rm e},j}\leq \lambda_h$ holds for each $j$, which implies that
\begin{align}
    1-\bar\lambda_{\ell}^k \leq \lambda_{\rm v}\leq 1-\bar\lambda_h^k.
\end{align}

Similarly to the hyperedge case shown in the previous section, one can derive the exact solution of the IET distribution for the node (shown in Supplementary Note~4). We then assume that $r_{\ell h},r_{h \ell}\ll 1$ to obtain an approximate analytical solution as
\begin{align}
    P_{\rm v}(\tau)\approx \frac{1}{\Omega_{\rm v}}\sum_{\vec n} \prod_{j=0}^k f_{\mu_j n_j} \lambda_{\rm v}(\vec \mu,\vec n)^2 \bar\lambda_{\rm v}(\vec \mu,\vec n)^{\tau-1},
    \label{eq:Ptau_v_approx}
\end{align}
where $\Omega_{\rm v}$ denotes the average event probability for the node and is given by
\begin{align}
    \Omega_{\rm v}= \sum_{\vec n} \prod_{j=0}^k f_{\mu_j n_j} \lambda_{\rm v}(\vec \mu,\vec n).
    \label{eq:Omega_v_define}
\end{align}
In Eq.~\eqref{eq:Ptau_v_approx}, $\vec n \equiv (n_0,n_1,\ldots,n_k)$ with $0\leq n_j\leq \mu_j$ for $j\in\{0,1,\ldots,k\}$, and  $\bar\lambda_{\rm v}\equiv 1-\lambda_{\rm v}$. Similar to the case of hyperedges, the average event probability for the node, $\Omega_{\rm v}$, decreases as $m$ increases under the AND rule [see Fig.~\ref{fig:omega}(c) for the case of $k=2$]. Under the LIN rule, $\Omega_{\rm v}$ is almost independent of $m$ for the parameter values used in Fig.~\ref{fig:omega}, whereas it slightly increases as $m$ increases [see Fig.~\ref{fig:omega}(d) for the case of $k=2$].

Equation~\eqref{eq:Ptau_v_approx} is a weighted sum of geometric distributions with different probabilities of $\{\lambda_{\rm v}(\vec \mu,\vec n)\}$ over all possible $\vec n$. For example, for a minimal non-trivial case in which the focal node belongs to two hyperedges of the same size, i.e., $k=2$ and $m_1=m_2=m$, the IET distribution, Eq.~\eqref{eq:Ptau_v_approx}, under the AND rule is a weighted sum of three geometric distributions (see Supplementary Note~5 for the derivation). Similar to the case of hyperedges, the weighted sum of geometric distributions implies that the IET distribution for nodes has a heavier tail than its corresponding Bernoulli process with the same mean event probability $\Omega_{\rm v}$.

\subsubsection*{Numerical results}\label{subsec:IET_simulation}

We perform numerical simulations to test the accuracy of our theory and assess the generated average event probabilities and IET distributions (see the Methods section for further details of the numerical procedure).

\begin{figure}[!t]
    \centering
    \includegraphics[width=\columnwidth]{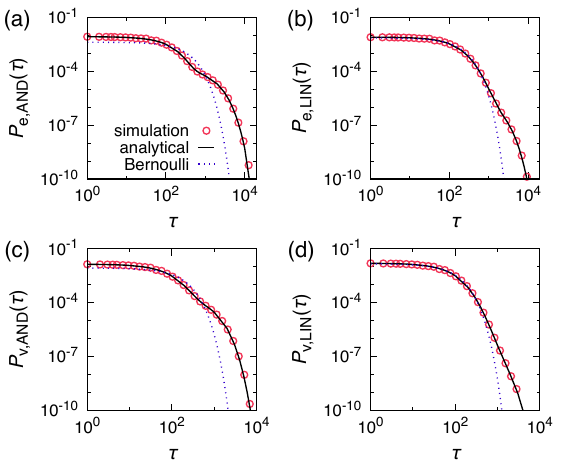}
    \caption{Interevent time distributions $P(\tau)$. (a) Hyperedge, AND rule. (b) Hyperedge, LIN rule. (c) Node, AND rule. (d) Node, LIN rule. For the numerical simulation (red circles), we reuse the event sequences with $m=3$ generated for Fig.~\ref{fig:omega}. For the analytical results (black solid lines), we numerically calculate Eqs.~(S12) and~(S39). For comparison to the Bernoulli processes, we also plot the geometric interevent time distributions 
    in Eq.~\eqref{eq:Ptau_e_avg} (blue dotted lines), with $\Omega_{\rm e}$ given by Eqs.~\eqref{eq:Omega_e_and_m} and~\eqref{eq:Omega_e_lin_m} for hyperedges, and with $\Omega_{\rm e}$ being replaced by $\Omega_{\rm v}$ in Eqs.~(S41) and~(S45) for nodes.
    }
    \label{fig:IET}
\end{figure}

Figure~\ref{fig:omega} shows that the numerical results for the average event probability are in good agreement with the analytical solutions. For the IET distributions, we only show the cases of $m=3$ in Fig.~\ref{fig:IET}. We again find that the numerical results are in good agreement with the analytical results. Figure~\ref{fig:IET} also suggests that the IET distributions for both hyperedges and nodes in our model (shown by the circles and lines) are longer-tailed than the IET distributions for the Bernoulli process with the same average event probability (shown by the dotted lines).

To quantify longer tails of the IET distributions for our models, we measure the coefficient of variation (CV), which is the standard deviation divided by the average. The CV of IETs is larger when the IET distribution has a heavier tail~\cite{Goh2008Burstiness, Kim2016Measuring}. Note that the Poisson process (in continuous-time) yields an exponential IET distribution, whose CV is equal to $1$; the geometric distribution, the discrete-time counterpart, with event probability $\lambda$ has CV $= \sqrt{1-\lambda}$. Using the approximate analytical solution of $P_{\rm e}(\tau)$ in Eq.~\eqref{eq:Ptau_e_approx}, we obtain
\begin{align}
    {\rm CV}_{\rm e} = \sqrt{\Omega_{\rm e}\left[
    \sum_{n=0}^m \frac{2f_{mn}}{\lambda_{\rm e}(m,n)}-1
    \right]-1}.
    \label{eq:cv_e_m}
\end{align}

\begin{figure}[!t]
    \centering
    \includegraphics[width=\columnwidth]{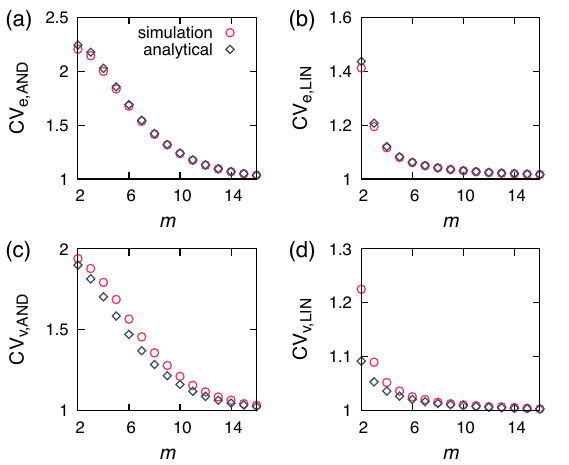}
    \caption{Coefficients of variation (CVs) of interevent times as a function of hyperedge size $m$. (a) Hyperedge, AND rule. (b) Hyperedge, LIN rule. (c) Node, AND rule. (d) Node, LIN rule. For the numerical simulation (red circles), we reuse the event sequences generated for Fig.~\ref{fig:omega}. We obtain the analytical results (dark blue diamonds) by numerically calculating Eqs.~\eqref{eq:cv_e_and},~\eqref{eq:cv_e_lin},~(S49),~(S50), and~(S51).
    }
    \label{fig:cv}
\end{figure}

For the AND rule, we use  Eq.~\eqref{eq:Ptau_e_and_m} to obtain
\begin{align}
    {\rm CV}_{\rm e,AND} = \sqrt{\left[p_h^m \lambda_h +(1-p_h^m)\lambda_{\ell}\right]\left[
    \tfrac{2p_h^m}{\lambda_h}+\tfrac{2(1-p_h^m)}{\lambda_{\ell}}-1
    \right]-1},
    \label{eq:cv_e_and}
\end{align}
which decreases with $m$ for $m\geq m_c$, where
\begin{align}
    m_c\equiv \frac{\ln \frac{1}{2}\left[1-\frac{\lambda_h \lambda_{\ell}}{2(\lambda_h -\lambda_{\ell})}\right]}{\ln p_h}.
\end{align}
For the same values of $r_{\ell h}$, $r_{h \ell}$, $\lambda_h$, and $\lambda_{\ell}$ as those used in Figs.~\ref{fig:omega} and \ref{fig:IET}, we obtain $m_c\approx 2.06$, implying that ${\rm CV}_{\rm e,AND}$ decreases with $m$ for any $m\ge 3$ and also likely so for $m\ge 2$. The numerical results shown in Fig.~\ref{fig:cv}(a) confirm these predictions. This decreasing trend is consistent with the fact that most events occur with the lower probability $\lambda_{\ell}$ if $m$ is large. In the limit of $m\to\infty$, our model is reduced to a Bernoulli process, resulting in a geometric IET distribution and ${\rm CV}_{\rm e,AND} = \sqrt{1-\lambda_{\ell}}$.

For the LIN rule, we use Eq.~\eqref{eq:Ptau_e_lin_m} to obtain
\begin{align}
    {\rm CV}_{\rm e,LIN} = \sqrt{(p_h\lambda_h + \bar p_h\lambda_{\ell})\left[
    \sum_{n=0}^m \frac{2f_{mn}}{\frac{n}{m}\lambda_h+(1-\frac{n}{m})\lambda_{\ell}}-1
    \right]-1}.
    \label{eq:cv_e_lin}
\end{align}
For the same values of $r_{\ell h}$, $r_{h \ell}$, $\lambda_h$, and $\lambda_{\ell}$ as in the AND case, we find that ${\rm CV}_{\rm e,LIN}$ decreases with $m$ for $m=2,\ldots, 16$ [see Fig.~\ref{fig:cv}(b)]. In the limit of $m\to\infty$, the most probable value of $m_h$, i.e., the number of nodes in the $h$ state, approaches $m/2$, such that an event occurs with the probability $(\lambda_h+\lambda_{\ell})/2$ at any time. Then, the event-generation process becomes a Bernoulli process, leading to the geometric IET distribution with ${\rm CV}_{\rm e,LIN} = \sqrt{1-(\lambda_h+\lambda_{\ell})/2}$.

One can also derive the CVs for the node in a similar manner (see Supplementary Note~6 for the derivation). Figures~\ref{fig:cv}(c) and (d) validate the theoretical expressions against numerical results. The same figures also show that the CV decreases with $m$ under both AND and LIN rules, which is qualitatively the same result as that for hyperedges shown in Figs.~\ref{fig:cv}(a) and (b).

\subsection*{Autocorrelation functions}\label{sec:ACF}

\subsubsection*{ACF for hyperedges}\label{subsec:ACF_hyper}

We derive the ACF for the time series of $e(t)$; we remind that $e(t)=1$ when an event occurs at time $t$, and $e(t) = 0$ otherwise. The ACF is defined as~\cite{Jo2024Temporal, Hartle2025Autocorrelation}
\begin{align}
    A_{\rm e}(t_{\rm d})\equiv \frac{ \langle e(t)e(t+t_{\rm d})\rangle- \langle e(t)\rangle^2}{ \langle e(t)^2\rangle- \langle e(t)\rangle^2},
    \label{eq:acf_e_define}
\end{align}
where $\langle\cdot\rangle$ is the time average, and $t_{\rm d}$ is the time lag. By definition, $A_{\rm e}(0)=1$. We consider $t_{\rm d}>0$ without loss of generality. We first note that
\begin{align}
    \langle e(t)\rangle=\Omega_{\rm e},\ \langle e(t)^2\rangle=\langle e(t)\rangle,
\label{eq:e(t)-relation}
\end{align}
where the second relation is due to the fact that $e(t) \in \{0, 1 \}$. We write
\begin{align}
    \langle e(t)e(t+t_{\rm d})\rangle = \langle e(t)\rangle R_{\rm e}(t_{\rm d}),
\label{eq:e(t)-R(t_d)}
\end{align}
where
\begin{align}
    R_{\rm e}(t_{\rm d})\equiv \Pr[e(t+t_{\rm d})=1|e(t)=1].
    \label{eq:Rtd_e_define}
\end{align}
By substituting Eqs.~\eqref{eq:e(t)-relation} and \eqref{eq:e(t)-R(t_d)} in Eq.~\eqref{eq:acf_e_define}, we obtain
\begin{align}
    A_{\rm e}(t_{\rm d})=\frac{ R_{\rm e}(t_{\rm d})- \Omega_{\rm e}}{1- \Omega_{\rm e}}.
    \label{eq:acf_e_define_simple}
\end{align}
To derive $R_{\rm e}(t_{\rm d})$ for any fixed $m$, we set $t=0$ without loss of generality in Eq.~\eqref{eq:Rtd_e_define} to obtain
\begin{align}
    R_{\rm e}(t_{\rm d}) =& \Pr[e(t_{\rm d})=1|e(0)=1]\notag\\
    =& \sum_{n,n'=0}^m 
    \Pr[e(t_{\rm d})=1|m_h(t_{\rm d})=n] \notag\\
    &\times \Pr[m_h(t_{\rm d})=n|m_h(0)=n'] \notag\\
    &\times \Pr[m_h(0)=n'|e(0)=1],
    \label{eq:Re_td_define}
\end{align}
where $m_h(t)$ is the number of $h$-state nodes in the hyperedge at time $t$. From Eqs.~\eqref{eq:Pr_1n} and~\eqref{eq:Pr_n1}, we obtain 
\begin{align}
    &\Pr[e(t_{\rm d})=1|m_h(t_{\rm d})=n] = \lambda_{\rm e}(m,n), \label{eq:Pr_1n_acf}\\
    &\Pr[m_h(0)=n'|e(0)=1]=\frac{f_{m n'}\lambda_{\rm e}(m,n')}{\Omega_{\rm e}}.
    \label{eq:Pr_n1_acf}
\end{align}
To compute $\Pr[m_h(t_{\rm d})=n|m_h(0)=n']$, we remind that $m_h(t)$ obeys the master equation given by Eq.~\eqref{eq:Pnt_master}, enabling us to write
\begin{align}
    \Pr[m_h(t_{\rm d})=n|m_h(0)=n'] = \left[W_{\rm e}^{t_{\rm d}}\right]_{nn'},
\label{eq:W_n|n'(t_d)}    
\end{align}
where the entries of matrix $W_{\rm e}$ are given by Eq.~\eqref{eq:Wenn'_define}. One can compactly write $R_{\rm e}(t_{\rm d})$ by a product of vectors and a matrix as
\begin{align}
    R_{\rm e}(t_{\rm d})= \vec\Lambda_{\rm e,f}^\top  W_{\rm e}^{t_{\rm d}} \vec\Lambda_{\rm e,i}.
    \label{eq:Re_td_matrix}
\end{align}
where the expressions of vectors $\vec\Lambda_{\rm e,f}$ and $\vec\Lambda_{\rm e,i}$ are given in Supplementary Note~2. In sum, by combining Eqs.~\eqref{eq:Pr_1n_acf}--\eqref{eq:W_n|n'(t_d)}, one can derive $R_{\rm e}(t_{\rm d})$ in Eq.~\eqref{eq:Re_td_define}, or equivalently Eq.~\eqref{eq:Re_td_matrix}, and hence $A_{\rm e}(t_{\rm d})$ in Eq.~\eqref{eq:acf_e_define_simple}. 

To demonstrate the derivation of the ACF for a hyperedge under the AND rule, we have derived the ACFs for $m=2$, $3$, and $4$ as follows (Supplementary Note~7 for the derivation):
\begin{align}
    A_{\rm e,AND}(t_{\rm d})=&\frac{(\lambda_h-\lambda_{\ell})^2 p_h^{2m}}{[p_h^m\lambda_h+(1-p_h^m)\lambda_{\ell}][1-p_h^m\lambda_h-(1-p_h^m)\lambda_{\ell}]} \notag\\
    &\times \sum_{n=1}^m {m\choose n}\left(\frac{\bar p_h}{p_h}\right)^n \eta^{nt_{\rm d}},
    \label{eq:acf_e_and_m}
\end{align}
where $\eta$ is given by Eq.~\eqref{eq:eta-def}. Under the LIN rule, we have obtained the following expression for $m=2$,~$3$, and~$4$ (Supplementary Note~8 for the derivation):
\begin{align}
    A_{\rm e,LIN}(t_{\rm d})=\frac{(\lambda_h-\lambda_{\ell})^2 p_h\bar p_h\eta^{t_{\rm d}}}{m(p_h\lambda_h+\bar p_h\lambda_{\ell})(1- p_h\lambda_h-\bar p_h\lambda_{\ell})}.
    \label{eq:acf_e_lin_m}
\end{align}

Respecting the fact that our model is reduced to a Bernoulli process when $\lambda_h=\lambda_{\ell}$, both Eqs.~\eqref{eq:acf_e_and_m} and~\eqref{eq:acf_e_lin_m} yield $A_{\rm e}(t_{\rm d})=\delta_{t_{\rm d},0}$ when $\lambda_h=\lambda_{\ell}$.

\subsubsection*{ACF for nodes}\label{subsec:ACF_node}

We have similarly derived the ACFs for nodes. The ACF is a mixture of exponential functions of $t_{\rm d}$ (see Supplementary Note~9). In the Bernoulli case (i.e., when $\lambda_h=\lambda_{\ell}$), we obtain $A_{\rm v}(t_{\rm d}) = \delta_{t_{\rm d},0}$ as expected.

\begin{figure}[!t]
    \centering
    \includegraphics[width=\columnwidth]{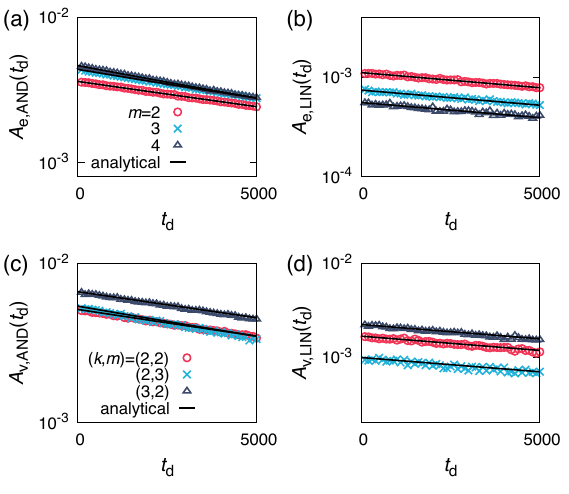}
    \caption{Autocorrelation functions $A$ as a function of time lag $t_{\rm d}$. (a) Hyperedge, AND rule. (b) Hyperedge, LIN rule. (c) Node, AND rule. (d) Node, LIN rule. 
    For the numerical simulation on the hyperedge (red circles for $m=2$, sky blue crosses for $m=3$, dark blue triangles for $m=4$), we reuse the event sequences generated for Fig.~\ref{fig:omega}. For the numerical simulation on the node with  $(k,m)=(2,2)$ (red circles) and $(2,3)$ (sky blue crosses), we reuse the event sequences generated for Fig.~\ref{fig:omega}. For the node with $(k,m)=(3,2)$ (dark blue triangles), we generate $10^3$ event sequences. Error bars are omitted because they are smaller than the symbols. To obtain the analytical results (black solid lines), we numerically calculate Eqs.~\eqref{eq:acf_e_and_m},~\eqref{eq:acf_e_lin_m},~(S97),~(S101),~(S108),~(S116),~(S119), and~(S123).
    }
    \label{fig:ACF}
\end{figure}

\subsubsection*{Numerical results}\label{subsec:ACF_simulation}

We perform numerical simulations to verify our analytical results (see the Methods section for further details of the numerical procedure). We compare the analytically and numerically obtained ACFs for a few values of $m$ and $k$ in Fig.~\ref{fig:ACF}. The numerical results are in good agreement with the analytical solutions.

\subsection*{Analysis of empirical hypergraphs}

We analyze publicly available data of temporal hypergraphs to compare their event rates, IET statistics, and ACFs with our theoretical predictions. We have downloaded all the 17 temporal hypergraphs from Austin R. Benson's website~\cite{Benson2025Austin}. We have selected six out of the 17 temporal hypergraphs. We exclude nine temporal hypergraphs because they show strong daily and/or weekly periodic behaviors in terms of the aggregate number of events; our model does not intend to explain such periodic behaviors. We also exclude two other temporal hypergraphs because they do not have enough hyperedges for the ACF analysis. Basic properties of the six temporal hypergraphs used are shown in Table~\ref{tab:data}. For more details of these hypergraphs, see Ref.~\cite{Benson2018Simplicial}. 

In the high and primary school hypergraphs, nodes represent individuals, and the event on the hyperedge indicates that individuals forming the hyperedge are physically close to each other at the same time. In the DBLP computer science bibliography (DBLP) hypergraph, nodes represent authors, and events represent joint publications. In the drug abuse warning network (DAWN) data, nodes are drugs. Events on hyperedges represent the set of drugs that are simultaneously used by patients before their visits to the emergency department. The NDC stands for the national drug code; each event on the hyperedge represents a drug. Nodes are either class labels applied to the drug (for the NDC-class hypergraph) or substances making up the drug (for the NDC-substance hypergraph). In these two temporal hypergraphs, the time of the event is the date when the drug was first marketed.

\begin{table}[!t]
\centering
\caption{Descriptive statistics of the six empirical temporal hypergraphs. We remind that $N$ is the number of nodes, $E$ is the number of hyperedges, and $T$ is the number of time points.}
\begin{tabular}{l|ccccc}
\hline
Hypergraph name & $N$ & $E$ & $T$ & Time unit \\
\hline
High school & 327 & 7,818 & 18,178 & 20 sec\\
Primary school & 242 & 12,704 & 5,845 & 20 sec\\
DBLP & 1,924,991 & 2,466,799 & 82 & 1 year\\
DAWN & 2,558 & 141,087 & 31 & 3 months\\
NDC-class & 1,161 & 1,088 & 42,997 & 1 day\\
NDC-substance & 5,311 & 9,906 & 42,997 & 1 day\\
\hline
\end{tabular}
\label{tab:data}
\end{table}

\begin{figure}[!t]
    \centering
    \includegraphics[width=\columnwidth]{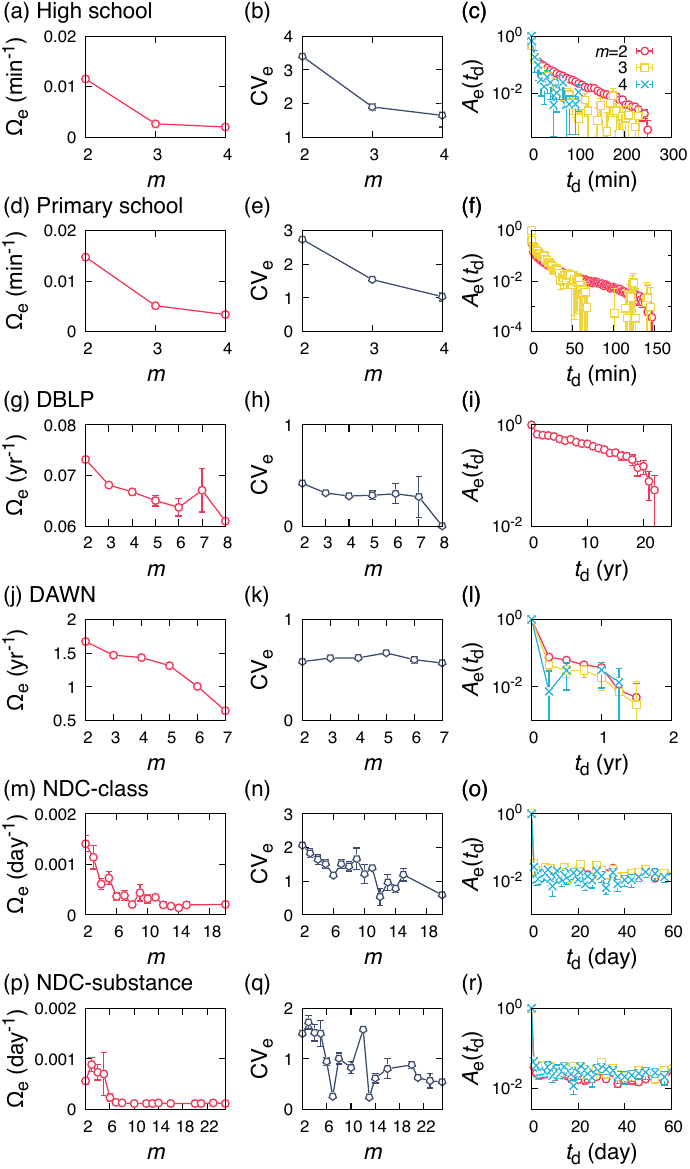}
    \caption{Results of empirical data analysis. We show the average event probability $\Omega_{\rm e}$ as a function of hyperedge size $m$ (a,~d,~g,~j,~m,~p), the coefficient of variation of interevent times (CV$_{\rm e}$) as a function of $m$ (b,~e,~h,~k,~n,~q), and the autocorrelation function as a function of time lag $A_{\rm e}(t_{\rm d})$ (c,~f,~i,~l,~o,~r), respectively, each for hyperedges. Each row of the figure corresponds to one empirical temporal hypergraph. In all panels, error bars stand for standard errors. Refer to Supplementary Note~10 for numbers of hyperedges analyzed in each panel.
    }
    \label{fig:data}
\end{figure}

We only focus on the statistical properties of hyperedges, but nodes. This is because there are too few nodes that have the same degree and the same size sequence of the hyperedges in each temporal hypergraph, hampering statistically reliable analyses. For each temporal hypergraph, we measure the average event probability, CV of IETs, and ACF as a function of $m$ (i.e., size of hyperedge). The average event probability is defined as the number of events on the hyperedge divided by $T$, the total time. For both average event probability and CV, we only consider the hyperedges with at least $5$ events. The ACF is defined by Eq.~\eqref{eq:acf_numeric}. We only calculate the ACF for the hyperedges with at least $20$ events. 

We show the average event probability, CV, and ACF in the left, middle, and right columns in Fig.~\ref{fig:data}. Each row of the figure corresponds to an empirical temporal hypergraph. In each panel, we show the average and standard error computed from all the qualified hyperedges of the given $m$. The figure indicates that, in a majority of hypergraphs, both the average event probability and CV tend to decrease with $m$. This result suggests that these temporal hypergraphs fit better the AND than LIN rule in terms of these two quantities. However, this last result does not exclude the possibility that the LIN rule is a plausible mechanism behind the empirical findings. For example, by studying the case in which $\lambda_h$ and $\lambda_{\ell}$ linearly decrease with $m$, we have found that the LIN rule can reproduce qualitatively the same behavior as that of the AND rule (see Supplementary Note~11).
The ACFs are overall exponentially decreasing with the time lag $t_{\rm d}$. This last trend qualitatively agrees with the behavior of our models (Fig.~\ref{fig:ACF}), including the case in which $\lambda_h$ and $\lambda_{\ell}$ depend on $m$ (see Supplementary Note~11). However, the dependence of the ACF on $m$ shown in the right column of Fig.~\ref{fig:data} is more complicated than that for our model. 

To test the significance of the empirical results, we adopt randomized reference models (RRMs)~\cite{Gauvin2022Randomized, Jo2020Bursttree} of the event sequence on each hyperedge. We find that temporal correlations due to the IET heterogeneity are significant, while the temporal correlation between IETs on the same hyperedge contributes negligibly to the ACF (Supplementary Note~12).

\section*{Discussion}\label{sec:conclusion}

Assuming that nodes obey a Markovian process on two (or a few) discrete states to explain non-Poissonian IET statistics is not a new idea~\cite{Malmgren2008Poissonian, Malmgren2009Universality, Karsai2012Universal}. One of the authors previously analyzed simple variants of such models for networks (i.e., dyadic interaction), showing that the IET distribution is more heavy-tailed than the Poissonian case~\cite{FonsecaDosReis2020Generative, FonsecaDosReis2022Metapopulation} and that the ACF shows non-trivial decay as a function of time lag~\cite{Hartle2025Autocorrelation}. The present analysis extends these results to the case of hypergraphs. The ACF results are qualitatively similar between hypergraphs and networks. Therefore, we attribute the ACF results to the node-state dynamics rather than to higher-order interaction. We have shown in Results section that the ACF decays according to a mixture of geometric decays, as opposed to the delta function expected from Poisson processes for event generation. Our analysis of empirical temporal hypergraph data confirmed a slow decay of the ACF, qualitatively consistent with our theory. In contrast, our IET results depend on the hyperedge size, $m$. Specifically, both the event rate and the spread of IETs (in terms of the CV) decrease as $m$ increases. We have confirmed that such dependencies on $m$ roughly hold true for a majority of empirical hypergraphs we analyzed. Therefore, we attribute the IET results to both the node-state dynamics and higher-order interaction, with the latter weakening the effect of the former. Future work includes further and larger-scale comparison between networks and hypergraphs, and inference of models from empirical hypergraph data, including examination of whether the AND or LIN rule better explains the data.

To indicate whether the AND rule, the LIN rule, or any other alternative fits better to empirical data, we would need statistical approaches. Such statistical work will provide us with quantitative insights into mechanisms generating time-stamped events in group-interaction settings, including whether our models or their variants are plausible enough for empirical data. We are not aware of a solid statistical framework even for network (and therefore simpler) versions of the present models~\cite{FonsecaDosReis2020Generative, Hartle2025Autocorrelation}. Methods for statistically fitting point-process models to generate neuronal spike trains with switching between two latent states~\cite{Chen2009Discrete, Tokdar2010Detection} and different methods for inferring the model parameters for time-stamped event sequences for a single individual~\cite{Malmgren2008Poissonian, Malmgren2009Universality} may be useful. Inference of our models is probably more difficult than that realized in these previous studies. This is because, in our models, a node is involved in multiple hyperedges (or edges) in general, and a hyperedge's events are determined by time-dependent states of multiple nodes.

Dynamical processes considered on temporal hypergraph models include spreading processes~\cite{Petri2018Simplicial, Chowdhary2021Simplicial, Zhao2022Effects, Han2024Probabilistic, Zino2024SusceptibleInfectedSusceptible, Mancastroppa2025Emerging, Shi2026Epidemic}, evolutionary games~\cite{Xu2023Higherorder}, and reaction-diffusion dynamics~\cite{Shi2025Turing}. Nevertheless, most of these studies remain numerical~\cite{Mancastroppa2025Emerging} or use HAD-type models while giving up non-Poissonian nature of event sequences at the expense of analytical tractability~\cite{Petri2018Simplicial, Han2024Probabilistic, Zino2024SusceptibleInfectedSusceptible, Shi2026Epidemic} (but see Ref.~\cite{Zhao2022Effects}). Because our models enable analytical approaches to a certain extent, we expect that our temporal hypergraph models help us to analytically examine dynamical processes on temporal hypergraphs. We remark that network versions of our model indeed enabled us to gain analytical insights into epidemic processes on temporal networks, such as the expressions of the epidemic threshold~\cite{Masuda2020Small, Liu2024Effects}. Obtaining clear insights into the effects of, e.g., hyperedge sizes and non-Poissonian event-time statistics on dynamical processes under group interaction via such an approach warrants future work.

\section*{Methods}\label{sec:method}

\subsection*{Numerical procedures}\label{subsec:simulation}

We numerically generate stochastic time-stamped event sequences on hypergraphs as follows. To examine the hyperedge, we use a single hyperedge composed of $m$ nodes. To examine the node, we use a node belonging to $k$ hyperedges of the same size $m$ by preparing $k(m-1)+1$ nodes. One of the $k(m-1)+1$ nodes is used as a focal node. The other $k(m-1)$ nodes are distinct neighbors of the focal node. To calculate the average event probabilities and IET distributions, we use $m\in\{2,\ldots,16\}$ for the hyperedge, and $k=2$ and $m\in\{2,\ldots,16\}$ for the node. To calculate the ACFs, we use $m=2,3,4$ for the hyperedge, and $(k,m)=(2,2)$, $(2,3)$, and $(3,2)$ for the node. At the initial time, $t=1$, each node is set to the $h$ or $\ell$ state with the stationary probability $p_h^*$ or $1-p_h^*$, respectively [Eq.~\eqref{eq:ph_rr}]. At each time $t \in \{1, 2, \ldots, T \}$, each hyperedge generates an event on it with probability $\lambda_{\rm e}$, which is given by Eq.~\eqref{eq:event_rate_and} for the AND rule and Eq.~\eqref{eq:event_rate_lin} for the LIN rule. Then, each node in the $h$ or $\ell$ state flips to the opposite state (i.e., $\ell$ or $h$ state, respectively) with probability $r_{h\ell}$ or $r_{\ell h}$, respectively. This procedure is repeated until $t=T$ to obtain $\{ e(1), \ldots, e(T) \}$ for the hyperedge and $\{ v(1), \ldots, v(T) \}$ for the node. We generate $10^3$ sequences of time-stamped events of length $T=10^7$ using $r_{\ell h}=5\times 10^{-5}$, $r_{h \ell}=2\times 10^{-5}$, $\lambda_h=10^{-2}$, and $\lambda_{\ell}=10^{-3}$.

To calculate the average event probability, we divide the total number of events generated on each hyperedge and node by $10^3T$. We obtain the IET distribution for each hyperedge and node by aggregating all IETs from the $10^3$ event sequences.

Next, we numerically calculate the ACF for each sequence of time-stamped events of length $T$ on the hyperedge, $\{ e(1), \ldots, e(T) \}$, as
\begin{align}
    \tilde A_{\rm e}(t_{\rm d})\equiv \frac{\frac{1}{T-t_{\rm d}} \sum_{t=1}^{T-t_{\rm d}} e(t)e(t+t_{\rm d})-\lambda_1\lambda_2}{\sigma_1\sigma_2},
    \label{eq:acf_numeric}
\end{align}
where $\lambda_1$ and $\sigma_1$ are the average and standard deviation of $\{e(1), \ldots, e(T-t_{\rm d})\}$, respectively; $\lambda_2$ and $\sigma_2$ are the average and standard deviation of $\{e(t_{\rm d}+1),\ldots, e(T)\}$, respectively. We average $\tilde A_{\rm e}(t_{\rm d})$ over the $10^3$ runs. We similarly compute $\tilde A_{\rm v}(t_{\rm d})$ for each run using the numerically produced event sequence on a node, $\{ v(1), \ldots, v(T) \}$, which we average over the $10^3$ runs.


%

\section*{Author contributions}
N.M. conceived the research concept. Under the supervision of N.M., H.-H.J performed analytical, numerical, and empirical analyses. H.-H.J. and N.M. drafted and revised the manuscript.

\section*{Competing interests}
The authors declare no competing interests.

\section*{Funding}
H.-H.J. acknowledges financial support by the National Research Foundation of Korea (NRF) grant funded by the Korea government (MSIT) (RS-2026-25476645).
N.M. acknowledges support by the NSF under Grant No.\,DMS-2204936, JSPS KAKENHI under Grants No.\,23H03414, No.\,24K14840, and No.\,24K030130, and by Japan Science and Technology Agency (JST) under Grant No.\,JPMJMS2021.

\section*{Data availability}
The empirical data that support the findings of this study are available in Austin R. Benson's website at \url{https://www.cs.cornell.edu/~arb/data/}. The source data in all main figures are in Supplementary Data.

\end{document}



\title{Supplementary Information on Modeling non-Poissonian temporal hypergraphs by Markovian node dynamics}

\author{Hang-Hyun Jo}
\email{h2jo@catholic.ac.kr}
\affiliation{Department of Physics, The Catholic University of Korea, Bucheon, Republic of Korea}

\author{Naoki Masuda}
\email{naokimas@umich.edu}
\affiliation{Gilbert S. Omenn Department of Computational Medicine and Bioinformatics, University of Michigan, Ann Arbor, MI, USA}
\affiliation{Department of Mathematics, University of Michigan, Ann Arbor, MI, USA}
\affiliation{Center for Computational Social Science, Kobe University, Kobe, Japan}


\date{\today}

\maketitle



\section{IET distributions in continuous time}\label{append:continuous_iet}

We analyze the continuous-time variant of the AND and LIN models, deriving IET distributions for hyperedges and nodes. We largely reuse the notations introduced 
in the main text, with the caveat that $r_{\ell h}$, $r_{h\ell}$, $\lambda_h$, and $\lambda_{\ell}$ are rates, not probabilities, in this Supplementary Note.

\subsection{IET distribution for hyperedges}

As one of the authors did in the previous study for pairwise-interaction networks~\cite{FonsecaDosReis2020Generative}, using Eq.~\eqref{eq:fmn} in the main text, we approximate the IET distribution for the hyperedge by
\begin{align}
    P_{\rm e}(\tau)\approx \frac{1}{\Omega_{\rm e}} \sum_{n=0}^m f_{mn} \lambda_{\rm e}(m,n)^2 e^{-\lambda_{\rm e}(m,n)\tau},
    \label{eq:Ptau_e_define}
\end{align}
where
\begin{align}
    \Omega_{\rm e} \equiv \sum_{n=0}^m f_{mn} \lambda_{\rm e}(m,n)
    \label{eq:Omega_e_define}
\end{align}
is the average event rate on the hyperedge. We have ignored IETs spanning more than one hyperedge state; for example, under the AND rule, we ignore IETs spanning both $Y_j = h$ and $Y_j = \ell$. In other words, we have assumed that $Y_j$ seldom changes between two consecutive events determining an IET, i.e., $ r_{h\ell} \ll \lambda_h$ and $ r_{\ell h} \ll \lambda_{\ell}$. Our result is not valid when these conditions are violated, e.g., when $\lambda_{\ell}=0$.

For the AND rule, Eqs.~\eqref{eq:Ptau_e_define} and \eqref{eq:Omega_e_define} reduce to
\begin{align}
    P_{\rm e,AND}(\tau)\approx\frac{1}{\Omega_{\rm e,AND}}\left[
    p_h^m\lambda_h^2 e^{-\lambda_h\tau}+(1-p_h^m)\lambda_{\ell}^2 e^{-\lambda_{\ell}\tau}\right]
    \label{eq:Ptau_e_and}
\end{align}
and
\begin{align}
    \Omega_{\rm e,AND}= p^m_h\lambda_h+ (1-p^m_h)\lambda_{\ell},
    \label{eq:Omega_e_and}
\end{align}
respectively. Equation~\eqref{eq:Ptau_e_and} is a weighted sum of two exponential distributions with different rates. The corresponding Poisson process with the same event rate, $\Omega_{\rm e,AND}$, produces an exponential IET distribution given by
\begin{align}
    P_{\rm e,AND}^{\rm Poi}(\tau)=\Omega_{\rm e,AND} e^{-\Omega_{\rm e,AND}\tau}.
    \label{eq:Ptau_e_and_Poi_cont}
\end{align}
Because $\lambda_{\ell} < \Omega_{\rm e, AND}<\lambda_h$, the IET distribution given by Eq.~\eqref{eq:Ptau_e_and} has a heavier tail than the corresponding Poisson case given by Eq.~\eqref{eq:Ptau_e_and_Poi_cont}~\cite{Yannaros1994Weibull, Feldmann1998Fitting, Masuda2022Gillespie}.

For the LIN rule, we obtain
\begin{align}
    P_{\rm e,LIN}(\tau)\approx\frac{1}{\Omega_{\rm e,LIN}} \sum_{n=0}^m f_{m n} \left[\frac{n}{m}\lambda_h +\left(1-\frac{n}{m}\right)\lambda_{\ell}\right]^2 e^{-\left[\frac{n}{m}\lambda_h +\left(1-\frac{n}{m}\right)\lambda_{\ell}\right]\tau},
    \label{eq:Ptau_e_lin}
\end{align}
and
\begin{align}
    \Omega_{\rm e,LIN}= p_h\lambda_h+\bar p_h\lambda_{\ell}.
    \label{eq:Omega_e_lin}
\end{align}
Equation~\eqref{eq:Ptau_e_lin} is a weighted sum of $m+1$ exponential distributions with rates linearly ranging from $\lambda_{\ell}$ to $\lambda_h$. Therefore, the IET distribution for the LIN model has a heavier tail than that for the Poisson process with the same mean event rate, $\Omega_{\rm e,LIN}$, whose IET distribution reads
\begin{align}
    P_{\rm e,LIN}^{\rm Poi}(\tau)=\Omega_{\rm e,LIN} e^{-\Omega_{\rm e,LIN}\tau}.
    \label{eq:Ptau_e_lin_Poi}
\end{align}

\subsection{IET distribution for nodes}

Let us derive the IET distribution for a node belonging to $k$ hyperedges of sizes $\vec m=(m_1,\ldots,m_k)$. We remind that $\vec m$ and $\vec m_h$ are respectively equivalent to $\vec \mu$ and $\vec \mu_h$ in describing the states of the focal node and its neighbors. The event rate on the node is the sum of event rates on the hyperedges containing the node:
\begin{align}
    \lambda_{\rm v}(\vec \mu,\vec \mu_h) = \sum_{j=1}^k \lambda_{{\rm e},j}(1+\mu_j,\mu_{0h}+\mu_{jh}),
    \label{eq:event_node_edge_cont}
\end{align}
where $\lambda_{{\rm e},j}$ is given by Eqs.~\eqref{eq:event_rate_and} and \eqref{eq:event_rate_lin} for the AND and LIN rules, respectively. One then obtains the IET distribution for the node as follows:
\begin{align}
    P_{\rm v}(\tau) \approx\frac{1}{\Omega_{\rm v}}\sum_{\vec n} \prod_{j=0}^k f_{\mu_j n_j} \lambda_{\rm v}(\vec \mu,\vec n)^2 e^{-\lambda_{\rm v}(\vec \mu,\vec n)\tau},
    \label{eq:Ptau_v_approx_cont}
\end{align}
where the average event rate on the node is given by
\begin{align}
    \Omega_{\rm v}= \sum_{\vec n} \prod_{j=0}^k f_{\mu_j n_j} \lambda_{\rm v}(\vec\mu,\vec n).
    \label{eq:Omega_v_define_cont}
\end{align}
Here $\vec n=(n_0,n_1,\ldots,n_k)$ with $0\leq n_j\leq \mu_j$ for $j\in\{0,1,\ldots,k\}$. 

\section{IET distribution for hyperedges as matrix products}\label{append:Ptau_e_matrix}

By combining Eqs.~\eqref{eq:Pr_1n},~\eqref{eq:Pr_0n},~\eqref{eq:Wenn'_define},~\eqref{eq:Pr_nt+1nt}, and~\eqref{eq:Pr_n1}, we can compactly write $P_{\rm e}(\tau)$ in Eq.~\eqref{eq:Ptau_e_discrete} as
\begin{align}
    P_{\rm e}(\tau) = \vec \Lambda_{\rm e,f}^\top (W_{\rm e}L_{\rm e})^{\tau-1}W_{\rm e} \vec \Lambda_{\rm e,i},
    \label{eq:Ptau_edge_matrix}
\end{align}
where
\begin{align}
    \vec \Lambda_{\rm e,f} \equiv \begin{pmatrix}
        \lambda_{\rm e}(m,0)\\
        \lambda_{\rm e}(m,1)\\
        \vdots\\
        \lambda_{\rm e}(m,m)
    \end{pmatrix},\label{eq:Lambda_e_define-1}
\end{align}
\begin{align}
    L_{\rm e} \equiv \begin{pmatrix}
        \bar\lambda_{\rm e}(m,0) & 0 & \cdots & 0\\
        0 & \bar\lambda_{\rm e}(m,1) & \cdots & 0\\
        \vdots &\vdots & &\vdots\\
        0 & 0 & \cdots & \bar\lambda_{\rm e}(m,m)
    \end{pmatrix},\label{eq:Lambda_e_define-2}
\end{align}
\begin{align}
    \vec \Lambda_{\rm e,i} \equiv \frac{1}{\Omega_{\rm e}}\begin{pmatrix}
        f_{m0}\lambda_{\rm e}(m,0)\\
        f_{m1}\lambda_{\rm e}(m,1)\\
        \vdots\\
        f_{mm}\lambda_{\rm e}(m,m)
    \end{pmatrix},
    \label{eq:Lambda_e_define-3}
\end{align}
and $^{\top}$ denotes the transposition.

To proceed further, we assume that $r_{\ell h},r_{h \ell}\ll 1$, implying that the transitions of nodes from the $h$ state to the $\ell$ state and vice versa are rare. This assumption leads to $W_{\rm e}\approx I$, where $I$ is the identity matrix. To show this, we note that matrix $W_{\rm e}$ can be diagonalized into
\begin{equation}
    W_{\rm e}=B_{\rm e}D_{\rm e}B_{\rm e}^{-1},
    \label{eq:We_BDB1}
\end{equation}
where $D_{\rm e}$ is a diagonal matrix consisting of eigenvalues of $W_{\rm e}$, i.e.,
\begin{align}
    D_{\rm e}=\begin{pmatrix}
        1 & 0 & \cdots & 0\\
        0 & \eta & \cdots & 0\\
        \vdots & \vdots &  & \vdots\\
        0 & 0 & \cdots & \eta^m
    \end{pmatrix},
    \label{eq:De_matrix}
\end{align}
and $\eta$ is given by Eq.~\eqref{eq:eta-def}. The entries of matrix $B_{\rm e}=(B_{{\rm e},nn'})$ are given by
\begin{align}
    B_{{\rm e},nn'}=\sum_{u=\max\{m-n-n',0\}}^{\min\{m-n,m-n'\}}(-1)^{m-n-u}{n'\choose m-n-u}{m-n'\choose u}\alpha^u,
    \label{eq:Be_element}
\end{align}
where $\alpha\equiv r_{h \ell} / r_{\ell h}$. Therefore, we obtain
\begin{align}
    [B_{\rm e}^{-1}]_{nn'}=\frac{B_{{\rm e},m-n,m-n'}}{(\alpha+1)^m}.
\end{align}
By the assumption that $r_{\ell h},r_{h \ell}\ll 1$, we find that $\eta=1-r_{\ell h}-r_{h \ell}\approx 1$, leading to $D_{\rm e}\approx I$. Thus, $W_{\rm e}\approx B_{\rm e}B_{\rm e}^{-1}=I$. Therefore, one can obtain the approximate analytical solution of the IET distribution as 
\begin{align}
    P_{\rm e}(\tau) &\approx \vec \Lambda_{\rm e,f}^\top 
    L_{\rm e}^{\tau-1}
    \vec \Lambda_{\rm e,i}\notag\\
    &= \frac{1}{\Omega_{\rm e}}\sum_{n=0}^m f_{mn}\lambda_{\rm e}(m,n)^2 \bar\lambda_{\rm e}(m,n)^{\tau-1}.
    \label{eq:Ptau_e_approx_append}
\end{align}

\section{Duration distribution for hyperedges}\label{append:duration}

Here we study the duration distribution, which can be studied within our modeling framework. To be precise, the duration of activity on a hyperedge is defined as the number of consecutive events separated by the non-events (i.e., $e(t)=0$), and it is denoted by $d$. Then, the duration distribution reads
\begin{align}
    P_{\rm e}(d)=&\Pr[e(d+1)=0, e(d)=1,\ldots, e(1)=1|e(0)=0] \notag\\
    =&\sum_{n_0,\ldots,n_{d+1}=0}^m \Pr[e(d+1)=0|m_h(d+1)=n_{d+1}] \left[\prod_{t=1}^{d} \Pr[m_h(t+1)=n_{t+1}|m_h(t)=n_t]
    \Pr[e(t)=1|m_h(t)=n_t] \right] \notag \\ 
    &\times \Pr[m_h(1)=n_1|m_h(0)=n_0] \Pr[m_h(0)=n_0|e(0)=0].
    \label{eq:Pd_e_discrete_duration}
\end{align}
Using matrix $W_{\rm e}$ whose elements are given by Eq.~\eqref{eq:Wenn'_define}, we write $P_{\rm e}(d)$ as
\begin{align}
    P_{\rm e}(d) = \vec{\bar \Lambda}_{\rm e,f}^\top (W_{\rm e}\bar L_{\rm e})^d W_{\rm e} \vec{\bar \Lambda}_{\rm e,i},
    \label{eq:Pd_edge_matrix_duration}
\end{align}
where
\begin{align}
    \vec{\bar \Lambda}_{\rm e,f} \equiv \begin{pmatrix}
        \bar\lambda_{\rm e}(m,0)\\
        \bar\lambda_{\rm e}(m,1)\\
        \vdots\\
        \bar\lambda_{\rm e}(m,m)
    \end{pmatrix},\label{eq:Lambda_e_define-1_duration}
\end{align}
\begin{align}
    \bar L_{\rm e} \equiv \begin{pmatrix}
        \lambda_{\rm e}(m,0) & 0 & \cdots & 0\\
        0 & \lambda_{\rm e}(m,1) & \cdots & 0\\
        \vdots &\vdots & &\vdots\\
        0 & 0 & \cdots & \lambda_{\rm e}(m,m)
    \end{pmatrix},\label{eq:Lambda_e_define-2_duration}
\end{align}
\begin{align}
    \vec{\bar\Lambda}_{\rm e,i} \equiv \frac{1}{\bar\Omega_{\rm e}}\begin{pmatrix}
        f_{m0}\bar\lambda_{\rm e}(m,0)\\
        f_{m1}\bar\lambda_{\rm e}(m,1)\\
        \vdots\\
        f_{mm}\bar\lambda_{\rm e}(m,m)
    \end{pmatrix}.
    \label{eq:Lambda_e_define-3_duration}
\end{align}
Here we have defined $\bar\Omega_{\rm e}\equiv 1-\Omega_{\rm e}$. To proceed further, we assume that $r_{\ell h},r_{h \ell}\ll 1$, leading to $W_{\rm e}\approx I$. Then, one can obtain the approximate analytical solution of the duration distribution as 
\begin{align}
    P_{\rm e}(d) &\approx \vec{\bar\Lambda}_{\rm e,f}^\top \bar L_{\rm e}^d \vec{\bar\Lambda}_{\rm e,i}\notag\\
    &= \frac{1}{\bar\Omega_{\rm e}}\sum_{n=0}^m f_{mn}\bar\lambda_{\rm e}(m,n)^2 \lambda_{\rm e}(m,n)^d.
    \label{eq:Ptau_e_approx_append_duration}
\end{align}
Equation~\eqref{eq:Ptau_e_approx_append_duration} is a weighted sum of geometric distributions with different non-event probabilities ranging from $\bar\lambda_{\rm e}(m,0)$ to $\bar\lambda_{\rm e}(m,m)$. Because $\bar\lambda_h<\bar\Omega_{\rm e}<\bar\lambda_{\ell}$ from Eq.~\eqref{eq:Omega_e}, our model has a heavier tail in the duration distribution than the corresponding Bernoulli process with the same mean non-event probability $\bar\Omega_{\rm e}$~\cite{Yannaros1994Weibull, Feldmann1998Fitting, Masuda2022Gillespie}, whose duration distribution is a geometric distribution given by
\begin{align}
    P_{\rm e}^{\rm B}(d)=\bar\Omega_{\rm e} (1-\bar\Omega_{\rm e})^d=(1-\Omega_{\rm e}) \Omega_{\rm e}^d.
    \label{eq:Ptau_e_avg_duration}
\end{align}

\begin{figure}[!t]
    \centering
    \includegraphics[width=0.5\textwidth]{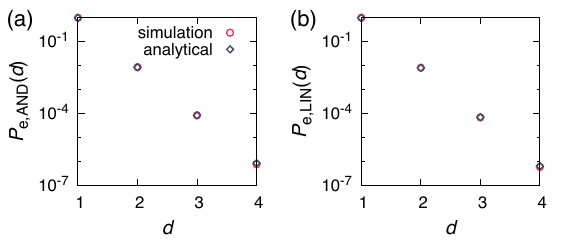}
    \caption{Duration distributions $P_{\rm e}(d)$ under AND (a) and LIN (b) rules. For the numerical simulation (circles), we reuse the event sequences with $m=3$ generated for Fig.~\ref{fig:omega}. For the analytical results (diamonds), we numerically calculate Eq.~\eqref{eq:Pd_edge_matrix_duration} using Eqs.~\eqref{eq:event_rate_and} and~\eqref{eq:event_rate_lin} for AND and LIN rules, respectively.
    }
    \label{fig:duration}
\end{figure}

Under the AND rule, we use Eq.~\eqref{eq:event_rate_and} to obtain
\begin{align}
    P_{\rm e,AND}(d)\approx \frac{1}{\bar\Omega_{\rm e,AND}}\left[
    p_h^m\bar\lambda_h^2 \lambda_h^d
    +(1-p_h^m)\bar\lambda_{\ell}^2 \lambda_{\ell}^d
    \right].
    \label{eq:Ptau_e_and_m_duration}
\end{align}
Under the LIN rule, we use Eq.~\eqref{eq:event_rate_lin} to obtain
\begin{align}
    P_{\rm e,LIN}(d)\approx \frac{1}{\bar\Omega_{\rm e,LIN}}
    \sum_{n=0}^m f_{mn}\left[\frac{n}{m}\bar\lambda_h+\left(1-\frac{n}{m}\right)\bar\lambda_{\ell}\right]^2  \left[\frac{n}{m}\lambda_h+\left(1-\frac{n}{m}\right)\lambda_{\ell}\right]^d.
    \label{eq:Ptau_e_lin_m_duration}
\end{align}

We carry out numerical simulations using the parameter values of $r_{\ell h}=5\times 10^{-5}$, $r_{h \ell}=2\times 10^{-5}$, $\lambda_h=10^{-2}$, and $\lambda_{\ell}=10^{-3}$, as in the main text. It turns out that the range of $d$ is limited only up to $4$ when $m=3$. Thus, the numerically obtained duration distributions looks like a simple exponential function, as shown in Fig.~\ref{fig:duration}.

\section{Derivation of the IET distribution for a node in the general case}\label{append:Ptau_v_general}

The IET distribution for the node, $P_{\rm v}(\tau)$, can be written as
\begin{align}
    P_{\rm v}(\tau)=&\Pr[v(\tau)=1, v(\tau-1)=0,\ldots, v(1)=0|v(0)=1] \notag\\
    =&\sum_{\vec n_0,\ldots,\vec n_{\tau}}
    \Pr[v(\tau)=1|\vec \mu_h(\tau)
    =\vec n_\tau]
    \left[\prod_{t=1}^{\tau-1} \Pr[\vec \mu_h(t+1)=\vec n_{t+1}|\vec \mu_h(t)=\vec n_t]
    \Pr[v(t)=0|\vec \mu_h(t)=\vec n_t] \right]\notag \\ 
    & \times \Pr[\vec \mu_h(1)=\vec n_1|\vec \mu_h(0)=\vec n_0] \Pr[\vec \mu_h(0)=\vec n_0|v(0)=1],\label{eq:Ptau_v_discrete}
\end{align}
where $\vec n_t=(n_{t,0},n_{t,1},\ldots,n_{t,k})$ with $0\leq n_{t,j}\leq \mu_j$ for $j\in\{0,1,\ldots,k\}$.

To compute Eq.~\eqref{eq:Ptau_v_discrete}, we first note that
\begin{align}
    \Pr[v(t)=1|\vec \mu_h(t)=\vec n_t] &= \lambda_{\rm v}(\vec \mu,\vec n_t), \label{eq:Pr_1n_node}\\
    \Pr[v(t)=0|\vec \mu_h(t)=\vec n_t] &= 1-\lambda_{\rm v}(\vec \mu,\vec n_t),
    \label{eq:Pr_0n_node}
\end{align}
where $\lambda_{\rm v}(\vec \mu,\vec n_t)$ is given by Eq.~\eqref{eq:event_node_edge}.

The conditional probability $\Pr[\vec \mu_h(t+1)=\vec n_{t+1}|\vec \mu_h(t)=\vec n_t]$ in Eq.~\eqref{eq:Ptau_v_discrete} is given by
\begin{align}
    \Pr[\vec \mu_h(t+1)=\vec n_{t+1}|\vec \mu_h(t)=\vec n_t]= W_{{\rm v},\vec n_{t+1}\vec n_t},
    \label{eq:Pr_nt+1nt_node}
\end{align}
where the transition probability matrix $W_{\rm v}=(W_{{\rm v},\vec n_{t+1}\vec n_t})$ is the Kronecker product of $k+1$ matrices for hyperedges:
\begin{align}
    W_{\rm v}=\bigotimes_{j=0}^k W_{{\rm e},j}.
    \label{eq:Wv_We_define}
\end{align}
Each $W_{{\rm e},j}$ is a $(\mu_j+1)\times (\mu_j+1)$ matrix whose entries are given by
\begin{align}
    \left[W_{{\rm e},j}\right]_{n_jn_j'} =\sum_{u=0}^{n_j'}\sum_{u'=0}^{\mu_j-n_j'} {n_j' \choose u}r_{h \ell}^u \bar r_{h \ell}^{n_j'-u} {\mu_j-n_j' \choose u'}r_{\ell h}^{u'} \bar r_{\ell h}^{\mu_j-n_j'-u'}\delta_{n_j,n_j'-u+u'}.
    \label{eq:Wej_nn'}
\end{align}
Note that Eq.~\eqref{eq:Wv_We_define} implies that
\begin{align}
    W_{{\rm v},\vec n\vec n'}=\prod_{j=0}^k \left[W_{{\rm e},j}\right]_{n_jn_j'}.
    \label{eq:Wv_nn'}
\end{align}

Finally, to compute $\Pr[\vec \mu_h(0)=\vec n_0|v(0)=1]$ in Eq.~\eqref{eq:Ptau_v_discrete}, we use the Bayes' theorem~\cite{Ross2014First} to obtain
\begin{align}
    \Pr[\vec\mu_h(0)=\vec n_0|v(0)=1] =& \frac{\Pr[v(0)=1|\vec \mu_h(0)=\vec n_0] \Pr[\vec\mu_h(0) = \vec n_0]}{\Pr[v(0)=1]}\notag\\  
    =& \frac{\lambda_{\rm v}(\vec \mu,\vec n_0) \prod_{j=0}^k f_{\mu_j n_{0,j}}}{\Omega_{\rm v}}.
    \label{eq:Pr_n1_node}
\end{align}
We have assumed the equilibrium for $\vec n_0$ at $t=0$ and therefore used
\begin{align}
    \Pr[\vec\mu_h(0) = \vec n_0] = \prod_{j=0}^k f_{\mu_j n_{0,j}}
    \label{eq:Pr_n0}
\end{align}
to derive the second equality in Eq.~\eqref{eq:Pr_n1_node}. We have also used $\Pr[v(0)=1]=\Omega_{\rm v}$ to derive the second equality of Eq.~\eqref{eq:Pr_n1_node} because both sides indicate the average event probability on the node [Eq.~\eqref{eq:Omega_v_define}]. 

By combining Eqs.~\eqref{eq:Pr_1n_node},~\eqref{eq:Pr_0n_node},~\eqref{eq:Pr_nt+1nt_node},~\eqref{eq:Wej_nn'},~\eqref{eq:Wv_nn'}, and~\eqref{eq:Pr_n1_node}, we compactly write $P_{\rm v}(\tau)$ in Eq.~\eqref{eq:Ptau_v_discrete} as
\begin{align}
     P_{\rm v}(\tau)= \vec \Lambda_{\rm v,f}^\top (W_{\rm v}L_{\rm v})^{\tau-1}W_{\rm v} \vec \Lambda_{\rm v,i},
     \label{eq:Ptau_v_matrix_append}
\end{align}
where $\vec\Lambda_{\rm v,f}$ denotes a column vector composed of $\Pr[v(\tau)=1|\vec \mu_h(\tau)=\vec n_\tau]$ [see Eq.~\eqref{eq:Pr_1n_node}] for all $\vec n_\tau$, implying that the dimension of $\vec\Lambda_{\rm v,f}$ is $\prod_{j=0}^k (\mu_j+1)$. Matrix $L_{\rm v}$ is a diagonal matrix of size $\prod_{j=0}^k (\mu_j+1)\times \prod_{j=0}^k (\mu_j+1)$, whose diagonal entries are $\Pr[v(t)=0|\vec \mu_h(t)=\vec n_t]$ [see Eq.~\eqref{eq:Pr_0n_node}] for all $\vec n_t$. Finally, $\vec\Lambda_{\rm v,i}$ is a column vector composed of $\Pr[\vec \mu_h(0)=\vec n_0|v(0)=1]$ [see Eq.~\eqref{eq:Pr_n1_node}] for all $\vec n_0$. The dimension of $\vec\Lambda_{\rm v,i}$ is the same as that of $\vec\Lambda_{\rm v,f}$. In sum, once the hyperedge-size vector $\vec m$ for a node and the rule for generating events on hyperedges (i.e., AND or LIN) are given, one can derive the exact solution of $P_{\rm v}(\tau)$ given by Eq.~\eqref{eq:Ptau_v_discrete}, or equivalently by Eq.~\eqref{eq:Ptau_v_matrix_append}. 

We then assume that $r_{\ell h},r_{h \ell}\ll 1$, leading to $W_{\rm v}\approx I$. By computing Eqs.~\eqref{eq:Pr_1n_node},~\eqref{eq:Pr_0n_node},~\eqref{eq:Wv_nn'}, and~\eqref{eq:Pr_n1_node}, one can derive the approximate analytical solution of $P_{\rm v}(\tau)$ shown in Eq.~\eqref{eq:Ptau_v_approx}.

\section{Derivation of the IET distribution for a node in a minimal nontrivial case}\label{append:Ptau_v}

\subsection{AND rule}\label{append:Ptau_v_and}

Under the AND rule, we consider a simple nontrivial case in which the focal node belongs to two hyperedges of the same size $m$, i.e., $k=2$ and $m_1=m_2=m$. It implies that $\vec\mu=(1,m-1,m-1)$. Using $\lambda_{\rm e}$ given by Eq.~\eqref{eq:event_rate_and}, we write $\lambda_{\rm v,AND}$ in Eq.~\eqref{eq:event_node_edge} as
\begin{align}
    \lambda_{\rm v,AND}(\vec \mu,\vec n) = 1-\left[\delta_{m,n_0+n_1}\bar\lambda_h + (1-\delta_{m,n_0+n_1})\bar\lambda_{\ell}\right]
    \left[\delta_{m,n_0+n_2}\bar\lambda_h + (1-\delta_{m,n_0+n_2})\bar\lambda_{\ell}\right],
    \label{eq:rate_v_and_k2m_simple}
\end{align}
where $n_0\in\{0,1\}$ and $n_1,n_2\in \{0,\ldots,m-1\}$. By substituting Eq.~\eqref{eq:rate_v_and_k2m_simple} in Eq.~\eqref{eq:Omega_v_define}, we obtain
\begin{align}
    \Omega_{\rm v,AND} &= 1- p_h\sum_{n_1,n_2=0}^{m-1} f_{m-1, n_1} f_{m-1, n_2}
    \left[\delta_{m,1+n_1}\bar\lambda_h + (1-\delta_{m,1+n_1})\bar\lambda_{\ell}\right]
    \left[\delta_{m,1+n_2}\bar\lambda_h + (1-\delta_{m,1+n_2})\bar\lambda_{\ell}\right]
    -\bar p_h \bar\lambda_{\ell}^2\notag\\
    &= 1- p_h \left[p_h^{m-1}\bar\lambda_h+(1-p_h^{m-1})\bar\lambda_{\ell} \right]^2
    -\bar p_h \bar\lambda_{\ell}^2\notag\\
    &= 1-p_h^{2m-1} \bar\lambda_h^2 
    -2p_h^m(1-p_h^{m-1}) \bar\lambda_h\bar\lambda_{\ell}
    -(1-2p_h^m+p_h^{2m-1}) \bar\lambda_{\ell}^2.
    \label{eq:Omega_v_and_k2m}
\end{align}
This result can be understood as follows: because each hyperedge can be either in the $h$ or $\ell$ state, we have three cases, i.e., (i) both hyperedges are in the $h$ state, (ii) one hyperedge is in the $h$ state and the other is in the $\ell$ state, and (iii) both hyperedges are in the $\ell$ state. In case (i), the event probability for the node, $\lambda_{\rm v,AND}$ in Eq.~\eqref{eq:event_node_edge}, is $1-\bar\lambda_h^2$. This case occurs with probability $p_h^{2m-1}$ because all $2m-1$ nodes should be in the $h$ state at the same time. In case (ii), we obtain $\lambda_{\rm v,AND}=1-\bar\lambda_h\bar\lambda_{\ell}$. This case occurs with probability $2p_h^m(1-p_h^{m-1})$. It is because all $m$ nodes in the hyperedge in the $h$ state should be in the $h$ state, while at least one of the $m-1$ nodes in the other hyperedge in the $\ell$ state (except for the focal node) should be in the $\ell$ state. In case (iii), we obtain $\lambda_{\rm v,AND}=1-\bar\lambda_{\ell}^2$. This case occurs with probability $1-2p_h^m+p_h^{2m-1}$. We also remark that $\Omega_{\rm v,AND}$ decreases with $m$. 

Under the assumption that $r_{\ell h},r_{h \ell}\ll 1$, it is straightforward to derive $P_{\rm v,AND}(\tau)$ from Eq.~\eqref{eq:Ptau_v_approx} as follows:
\begin{align}
    P_{\rm v,AND}(\tau)\approx \frac{
    p_h^{2m-1}(1-\bar\lambda_h^2)^2(\bar\lambda_h^2)^{\tau-1}
    + 2p_h^m(1-p_h^{m-1})(1-\bar\lambda_h \bar\lambda_{\ell})^2(\bar\lambda_h \bar\lambda_{\ell})^{\tau-1}
    + (1-2p_h^m+p_h^{2m-1}) (1-\bar\lambda_{\ell}^2)^2(\bar\lambda_{\ell}^2)^{\tau-1}
    }{\Omega_{\rm v,AND}}.
    \label{eq:Ptau_v_and_k2m}
\end{align}
Equation~\eqref{eq:Ptau_v_and_k2m} is a weighted sum of three geometric distributions whose probabilities are $1-\bar\lambda_h^2$, $1-\bar\lambda_h\bar\lambda_{\ell}$, and $1-\bar\lambda_{\ell}^2$. Because $1-\bar\lambda_{\ell}^2 <\Omega_{\rm v,AND} <1-\bar\lambda_h^2$, the IET distribution given by Eq.~\eqref{eq:Ptau_v_and_k2m} has a heavier tail than that for the Bernoulli process with the same event probability, i.e.,
\begin{align}
    P_{\rm v,AND}^{\rm B}(\tau)=\Omega_{\rm v,AND} (1-\Omega_{\rm v,AND})^{\tau-1}.
    \label{eq:Ptau_v_and_Poi}
\end{align}
In the limit of $m\to \infty$, 
the dynamics of the node activity converges to the Bernoulli process with event probability $\Omega_{\rm v,AND} = 1-\bar\lambda_{\ell}^2$, implying that
Eq.~\eqref{eq:Ptau_v_and_k2m} converges to the geometric distribution given by Eq.~\eqref{eq:Ptau_v_and_Poi}.

\subsection{LIN rule}\label{append:Ptau_v_lin}

Under the LIN rule, we consider the same simple case as that considered for the AND rule; 
the focal node belongs to two hyperedges of the same size $m$, i.e., $k=2$ and $m_1=m_2=m$, implying that $\vec\mu=(1,m-1,m-1)$. Then, $\lambda_{\rm v,LIN}$ given by Eq.~\eqref{eq:event_node_edge} reads
\begin{align}
    \lambda_{\rm v,LIN}(\vec \mu,\vec n) = 1-\left[\frac{n_0+n_1}{m}\bar\lambda_h + \left(1-\frac{n_0+n_1}{m}\right)\bar\lambda_{\ell}\right]\left[\frac{n_0+n_2}{m}\bar\lambda_h + \left(1-\frac{n_0+n_2}{m}\right)\bar\lambda_{\ell}\right],
    \label{eq:rate_v_lin_k2m_simple}
\end{align}
where $n_0\in\{0,1\}$ and $n_1,n_2\in \{0,\ldots,m-1\}$. By substituting Eq.~\eqref{eq:rate_v_lin_k2m_simple} in Eq.~\eqref{eq:Omega_v_define}, we obtain
\begin{align}
    \Omega_{\rm v,LIN} &= 1- p_h\sum_{n_1,n_2=0}^{m-1} f_{m-1, n_1} f_{m-1, n_2}\left[\frac{1+n_1}{m}\bar\lambda_h + \left(1-\frac{1+n_1}{m}\right)\bar\lambda_{\ell}\right]\left[\frac{1+n_2}{m}\bar\lambda_h + \left(1-\frac{1+n_2}{m}\right)\bar\lambda_{\ell}\right]\notag\\
    &- \bar p_h\sum_{n_1,n_2=0}^{m-1} f_{m-1, n_1} f_{m-1, n_2}\left[\frac{n_1}{m}\bar\lambda_h + \left(1-\frac{n_1}{m}\right)\bar\lambda_{\ell}\right]\left[\frac{n_2}{m}\bar\lambda_h + \left(1-\frac{n_2}{m}\right)\bar\lambda_{\ell}\right]\notag\\
    & =1 -p_h\left[\frac{mp_h+\bar p_h}{m}\bar\lambda_h +
    \frac{(m-1)\bar p_h}{m}\bar\lambda_{\ell} \right]^2
    -\bar p_h\left[\frac{(m-1) p_h}{m}\bar\lambda_h +
    \frac{m\bar p_h+p_h}{m}\bar\lambda_{\ell} \right]^2\notag\\
    &= 1 - (p_h\bar \lambda_h+\bar p_h\bar \lambda_{\ell})^2-\tfrac{1}{m^2}p_h\bar p_h (\lambda_h-\lambda_{\ell})^2.
    \label{eq:Omega_v_lin_k2m}
\end{align}
We remark that $\Omega_{\rm v,LIN}$ increases with $m$. 

For the IET distribution, we show the special case with $m=3$: Under the assumption that $r_{\ell h},r_{h \ell}\ll 1$, we obtain the approximate IET distribution as
\begin{align}
    P_{\rm v,LIN}(\tau)\approx &\frac{1}{\Omega_{\rm v,LIN}}\left\{
    p_h^5 (1-\bar\lambda_h^2)^2 (\bar\lambda_h^2)^{\tau-1}
    \right. \notag\\ & \left. 
    +\tfrac{4}{3} p_h^4 \bar p_h [1-\bar\lambda_h (2\bar\lambda_h + \bar\lambda_{\ell})]^2[\bar\lambda_h (2\bar\lambda_h + \bar\lambda_{\ell})]^{\tau-1}
    \right. \notag\\ & \left. 
    +\tfrac{2}{3} p_h^3 \bar p_h^2 [1-\bar\lambda_h (\bar\lambda_h + 2\bar\lambda_{\ell})]^2[\bar\lambda_h (\bar\lambda_h + 2\bar\lambda_{\ell})]^{\tau-1}
    \right. \notag\\ & \left.  
    +\tfrac{1}{9} p_h^3 \bar p_h(3\bar p_h+1)[1-(2\bar\lambda_h+\bar\lambda_{\ell})^2]^2 [(2\bar\lambda_h+\bar\lambda_{\ell})^2]^{\tau-1}
    \right. \notag\\ & \left. 
    +\tfrac{4}{9} p_h^2 \bar p_h^2 [1-(2\bar\lambda_h+\bar\lambda_{\ell})(\bar\lambda_h+2\bar\lambda_{\ell})]^2[(2\bar\lambda_h+\bar\lambda_{\ell})(\bar\lambda_h+2\bar\lambda_{\ell})]^{\tau-1}
    \right. \notag\\ & \left. 
    +\tfrac{1}{9} p_h \bar p_h^3(3 p_h+1) [1-(\bar\lambda_h+2\bar\lambda_{\ell})^2]^2[(\bar\lambda_h+2\bar\lambda_{\ell})^2]^{\tau-1}
    \right. \notag\\ & \left.
    +\tfrac{2}{3} p_h^2 \bar p_h^3 [1-(2\bar\lambda_h+\bar\lambda_{\ell})\bar\lambda_{\ell}]^2[(2\bar\lambda_h+\bar\lambda_{\ell})\bar\lambda_{\ell}]^{\tau-1}
    \right. \notag\\ & \left. 
    +\tfrac{4}{3} p_h \bar p_h^4 [1-(\bar\lambda_h+2\bar\lambda_{\ell})\bar\lambda_{\ell}]^2[(\bar\lambda_h+2\bar\lambda_{\ell})\bar\lambda_{\ell}]^{\tau-1}
    \right. \notag\\ & \left. 
    +\bar p_h^5 (1-\bar\lambda_{\ell}^2)^2 (\bar\lambda_{\ell}^2)^{\tau-1}
    \right\},\label{eq:Ptau_v_lin_k2m3}
\end{align}
where
\begin{align}
    \Omega_{\rm v,LIN} = 1 - (p_h\bar \lambda_h+\bar p_h\bar \lambda_{\ell})^2-\tfrac{1}{9}p_h\bar p_h (\lambda_h-\lambda_{\ell})^2.
    \label{eq:Omega_v_lin_k2m3}
\end{align}
Equation~\eqref{eq:Ptau_v_lin_k2m3} is a weighted sum of nine geometric distributions with different event probabilities. Because $1-\bar\lambda_{\ell}^2< \Omega_{\rm v,LIN}<1-\bar\lambda_h^2$, the IET distribution given by Eq.~\eqref{eq:Ptau_v_lin_k2m3} has a heavier tail than that of the Bernoulli process with the same event probability, i.e.,
\begin{align}
    P_{\rm v,LIN}^{\rm B}(\tau)=\Omega_{\rm v,LIN} (1-\Omega_{\rm v,LIN})^{\tau-1}.
    \label{eq:Ptau_v_lin_Poi}
\end{align}
In the limit of $m\to \infty$, the dynamics of the node activity converges to a Bernoulli process with event probability $\Omega_{\rm v,LIN} = 1-(1-\Omega_{\rm e,LIN})^2$, implying that Eq.~\eqref{eq:Ptau_v_lin_k2m3} converges to the geometric distribution.

\section{Derivation of CV for a node}\label{append:cv_node}

We compute the CVs for the node in a similar manner as in the main text. Using the approximate analytical solution of $P_{\rm v}(\tau)$ given by Eq.~\eqref{eq:Ptau_v_approx}, we obtain
\begin{align}
    {\rm CV}_{\rm v} = \sqrt{\Omega_{\rm v}\left[
    \sum_{\vec n} \frac{2\prod_{j=0}^k f_{\mu_j n_j}}{\lambda_{\rm v}(\vec \mu, \vec n)}-1
    \right]-1}.
    \label{eq:cv_v_m}
\end{align}
For the AND rule with $k=2$ and $m_1=m_2=m$, we use Eq.~\eqref{eq:Ptau_v_and_k2m} to obtain
\begin{align}
    {\rm CV}_{\rm v,AND} = \sqrt{\Omega_{\rm v,AND}\left[
    2\left(\frac{p_h^{2m-1}}{1-\bar\lambda_h^2}+\frac{2p_h^m(1-p_h^{m-1})}{1-\bar\lambda_h\bar\lambda_{\ell}}+\frac{1-2p_h^m+p_h^{2m-1}}{1-\bar\lambda_{\ell}^2}\right)-1
    \right]-1},
    \label{eq:cv_v_and_k2m}
\end{align}
where $\Omega_{\rm v,AND}$ is given by Eq.~\eqref{eq:Omega_v_and_k2m}. 

One can also compute the CV under the LIN rule, i.e., ${\rm CV}_{\rm v,LIN}$, for $k=2$ and $m_1=m_2=m$ by substituting 
\begin{align}
    \lambda_{\rm v}(\vec \mu,\vec n) = 1-\left[\frac{n_0+n_1}{m}\bar\lambda_h + \left(1-\frac{n_0+n_1}{m}\right)\bar\lambda_{\ell}\right]\left[\frac{n_0+n_2}{m}\bar\lambda_h + \left(1-\frac{n_0+n_2}{m}\right)\bar\lambda_{\ell}\right]
    \label{eq:rate_v_lin_k2m}
\end{align}
in Eq.~\eqref{eq:cv_v_m}.
 
\section{Derivation of ACF for a hyperedge under the AND rule}\label{append:acf_edge_and}

To demonstrate the derivation of the ACF for a hyperedge under the AND rule, we first consider the simplest case of $m=2$, i.e., an edge in a network (as opposed to hypergraph). By computing Eqs.~\eqref{eq:Omega_e_and_m}, \eqref{eq:Lambda_e_define-1}, and~\eqref{eq:Lambda_e_define-3}, one obtains
\begin{align}
    \Omega_{\rm e,AND} = p_h^2\lambda_h+(1-p_h^2)\lambda_{\ell},
    \label{eq:Omega_e_and_m2}
\end{align}
\begin{align}
\vec \Lambda_{\rm e,f} = \begin{pmatrix}
        \lambda_{\ell}\\
        \lambda_{\ell}\\
        \lambda_h
    \end{pmatrix},
    \label{eq:Lambda_e_and_m2}
\end{align}
\begin{align}
    \vec \Lambda_{\rm e,i} = \frac{1}{\Omega_{\rm e,AND}}
    \renewcommand{\arraystretch}{1.2}
    \begin{pmatrix}
        \bar p_h^2\lambda_{\ell}\\
        2p_h\bar p_h\lambda_{\ell}\\
        p_h^2\lambda_h
    \end{pmatrix}.
    \label{eq:Lambda_e_and_m2}
\end{align}
The transition probability matrix given by Eq.~\eqref{eq:Wenn'_define} reads
\begin{align}
    W_{\rm e}=\begin{pmatrix} \bar r_{\ell h}^2 & r_{h \ell} \bar r_{\ell h} & r_{h \ell}^2 \\
    2r_{\ell h}\bar r_{\ell h} & r_{h \ell}r_{\ell h}+\bar r_{h \ell}\bar r_{\ell h} & 2r_{h \ell}\bar r_{h \ell}\\
    r_{\ell h}^2 & \bar r_{h \ell} r_{\ell h} & \bar r_{h \ell}^2
    \end{pmatrix},\label{eq:W_m2}
\end{align}
leading to
\begin{equation}
    B_{\rm e} = \begin{pmatrix}
        \alpha^2 & -\alpha & 1\\
        2\alpha & \alpha-1 & -2\\
        1 & 1 & 1
    \end{pmatrix}
    \label{eq:Be_De_m2-1}
\end{equation}
and
\begin{equation}
    D_{\rm e} = \begin{pmatrix}
        1 & 0 & 0\\
        0 & \eta & 0\\
        0 & 0 & \eta^2
    \end{pmatrix}.
    \label{eq:Be_De_m2-2}
\end{equation}
By substituting Eqs.~\eqref{eq:Omega_e_and_m2}--\eqref{eq:Be_De_m2-2} in Eq.~\eqref{eq:Re_td_matrix}, we derive $R_{\rm e}(t_{\rm d})$ and then $A_{\rm e}(t_{\rm d})$ as follows:
\begin{align}
    A_{\rm e,AND}(t_{\rm d})=\frac{(\lambda_h-\lambda_{\ell})^2p_h^4 (2\alpha\eta^{t_{\rm d}}+ \alpha^2\eta^{2t_{\rm d}})}{[p_h^2\lambda_h+(1-p_h^2)\lambda_{\ell}][1-p_h^2\lambda_h-(1-p_h^2)\lambda_{\ell}]}.
    \label{eq:acf_e_and_m2}
\end{align}

For $m=3$, one obtains
\begin{align}
    \Omega_{\rm e,AND} = p_h^3\lambda_h +(1-p_h^3)\lambda_{\ell},
    \label{eq:Omega_e_and_m3}
\end{align}
\begin{align}
    \vec \Lambda_{\rm e,f} = \begin{pmatrix}
        \lambda_{\ell}\\
        \lambda_{\ell}\\
        \lambda_{\ell}\\
        \lambda_h
    \end{pmatrix},
    \label{eq:Lambda_e_and_m3}
\end{align}
\begin{align}
    \vec \Lambda_{\rm e,i} = \frac{1}{\Omega_{\rm e,AND}}
    \renewcommand{\arraystretch}{1.2}
    \begin{pmatrix}
        \bar p_h^3\lambda_{\ell}\\
        3p_h\bar p_h^2\lambda_{\ell}\\
        3p_h^2\bar p_h\lambda_{\ell}\\
        p_h^3\lambda_h 
    \end{pmatrix},
    \label{eq:Lambda_e_and_m3}
\end{align}
\begin{align}
    B_{\rm e} = \begin{pmatrix}
        \alpha^3 & -\alpha^2 & \alpha & -1\\
        3\alpha^2 & \alpha^2-2\alpha & 1-2\alpha & 3\\
        3\alpha & 2\alpha-1 & \alpha-2 & -3\\
        1 & 1 & 1 & 1
    \end{pmatrix},
    \label{eq:Be_De_m3-1}
\end{align}
\begin{align}
    D_{\rm e} = \begin{pmatrix}
        1 & 0 & 0& 0\\
        0 & \eta & 0& 0\\
        0 & 0 & \eta^2& 0\\
        0 & 0 & 0 & \eta^3
    \end{pmatrix},
    \label{eq:Be_De_m3-2}
\end{align}
leading to 
\begin{align}
    A_{\rm e,AND}(t_{\rm d})=\frac{(\lambda_h-\lambda_{\ell})^2p_h^6 (3\alpha\eta^{t_{\rm d}}+3\alpha^2\eta^{2t_{\rm d}} +\alpha^3\eta^{3t_{\rm d}})}{[p_h^3\lambda_h +(1-p_h^3)\lambda_{\ell}][1-p_h^3\lambda_h-(1-p_h^3)\lambda_{\ell}]}.
    \label{eq:acf_e_and_m3}
\end{align}

For $m=4$, one obtains
\begin{align}
    \Omega_{\rm e,AND} &= p_h^4\lambda_h+(1-p_h^4)\lambda_{\ell},
    \label{eq:Omega_e_and_m4}
\end{align}
\begin{align}
\vec \Lambda_{\rm e,f} = \begin{pmatrix}
        \lambda_{\ell}\\
        \lambda_{\ell}\\
        \lambda_{\ell}\\
        \lambda_{\ell}\\
        \lambda_h
    \end{pmatrix},
    \label{eq:Lambda_e_and_m4}
\end{align}
\begin{align}
    \vec \Lambda_{\rm e,i} = \frac{1}{\Omega_{\rm e,AND}}
    \renewcommand{\arraystretch}{1.2}
    \begin{pmatrix}
        \bar p_h^4\lambda_{\ell}\\
        4p_h\bar p_h^3\lambda_{\ell}\\
        6p_h^2\bar p_h^2\lambda_{\ell}\\
        4p_h^3\bar p_h\lambda_{\ell}\\
        p_h^4\lambda_h 
    \end{pmatrix},
    \label{eq:Lambda_e_and_m4}
\end{align}
\begin{align}
    B_{\rm e} = \begin{pmatrix}
        \alpha^4 & -\alpha^3 & \alpha^2 & -\alpha & 1\\
        4\alpha^3 & \alpha^3-3\alpha^2 & -2\alpha^2+2\alpha & 3\alpha-1 & -4\\
        6\alpha^2 & 3\alpha^2-3\alpha & \alpha^2-4\alpha+1 & -3\alpha+3 & 6\\
        4\alpha & 3\alpha-1 & 2\alpha-2 & \alpha-3 & -4\\
        1 & 1 & 1 & 1 & 1
    \end{pmatrix},
    \label{eq:Be_De_m4-1}
\end{align}
\begin{align}
    D_{\rm e} = \begin{pmatrix}
        1 & 0 & 0& 0& 0\\
        0 & \eta & 0& 0& 0\\
        0 & 0 & \eta^2& 0& 0\\
        0 & 0 & 0 & \eta^3& 0\\
        0 & 0 & 0 & 0 & \eta^4
    \end{pmatrix},
    \label{eq:Be_De_m4-2}
\end{align}
leading to 
\begin{align}
    A_{\rm e,AND}(t_{\rm d})=\frac{(\lambda_h-\lambda_{\ell})^2p_h^8 (
    4\alpha\eta^{t_{\rm d}}
    +6\alpha^2\eta^{2t_{\rm d}} 
    +4\alpha^2\eta^{3t_{\rm d}} 
    +\alpha^4\eta^{4t_{\rm d}})}{[p_h^4\lambda_h +(1-p_h^4)\lambda_{\ell}][1- p_h^4\lambda_h-(1-p_h^4)\lambda_{\ell}]}.
    \label{eq:acf_e_and_m4}
\end{align}
These results for $A_{\rm e,AND}(t_{\rm d})$ with $m=2,3,4$ can be summarized into Eq.~\eqref{eq:acf_e_and_m}.

\section{Derivation of ACF for a hyperedge under the LIN rule}\label{append:acf_edge_lin}

Under the LIN rule, for the same minimal case with $m=2$ as in~\ref{append:acf_edge_and}, we compute Eqs.~\eqref{eq:Lambda_e_define-1} and~\eqref{eq:Lambda_e_define-3} to obtain
\begin{align}
    \vec \Lambda_{\rm e,f} = \begin{pmatrix}
        \lambda_{\ell}\\
        \tfrac{1}{2}(\lambda_{\ell}+\lambda_h)\\
        \lambda_h
    \end{pmatrix},
    \label{eq:Lambda_e_and_m2-1}
\end{align}
\begin{align}
    \vec \Lambda_{\rm e,i} = \frac{1}{\Omega_{\rm e,LIN}}\begin{pmatrix}
        \bar p_h^2\lambda_{\ell}\\
        p_h\bar p_h(\lambda_{\ell}+\lambda_h)\\
        p_h^2\lambda_h 
    \end{pmatrix},
    \label{eq:Lambda_e_and_m2-2}
\end{align}
where $\Omega_{\rm e,LIN} = p_h\lambda_h+\bar p_h\lambda_{\ell}$ in Eq.~\eqref{eq:Omega_e_lin_m}.
Using $B_{\rm e}$ and $D_{\rm e}$ given by Eqs.~\eqref{eq:Be_De_m2-1} and \eqref{eq:Be_De_m2-2}, respectively, we obtain 
\begin{align}
    A_{\rm e,LIN}(t_{\rm d})=\frac{(\lambda_h-\lambda_{\ell})^2 p_h\bar p_h\eta^{t_{\rm d}}}{2(p_h\lambda_h+\bar p_h\lambda_{\ell})(1-p_h\lambda_h-\bar p_h\lambda_{\ell})}.
    \label{eq:acf_e_lin_m2}
\end{align}

For $m=3$, we obtain
\begin{align}
    \vec \Lambda_{\rm e,f} = 
    \renewcommand{\arraystretch}{1.2}
    \begin{pmatrix}
        \lambda_{\ell}\\
        \tfrac{1}{3}(2\lambda_{\ell}+\lambda_h)\\
        \tfrac{1}{3}(\lambda_{\ell}+2\lambda_h)\\
        \lambda_h
    \end{pmatrix}
    \label{eq:Lambda_e_and_m3-1}
\end{align}
and
\begin{align}
    \vec \Lambda_{\rm e,i} = \frac{1}{\Omega_{\rm e,LIN}}
    \renewcommand{\arraystretch}{1.2}
    \begin{pmatrix}
        \bar p_h^3\lambda_{\ell}\\
        p_h\bar p_h^2(2\lambda_{\ell}+\lambda_h)\\
        p_h^2\bar p_h(\lambda_{\ell}+2\lambda_h)\\
        p_h^3\lambda_h 
    \end{pmatrix}.
    \label{eq:Lambda_e_and_m3-2}
\end{align}
Using $B_{\rm e}$ and $D_{\rm e}$ given by Eqs.~\eqref{eq:Be_De_m3-1} and \eqref{eq:Be_De_m3-2}, respectively, we obtain 
\begin{align}
    A_{\rm e,LIN}(t_{\rm d})=\frac{(\lambda_h-\lambda_{\ell})^2 p_h\bar p_h\eta^{t_{\rm d}}}{3(p_h\lambda_h+\bar p_h\lambda_{\ell})(1-p_h\lambda_h-\bar p_h\lambda_{\ell})}.
    \label{eq:acf_e_lin_m3}
\end{align}

For $m=4$, we obtain
\begin{align}
    \vec \Lambda_{\rm e,f} = 
    \renewcommand{\arraystretch}{1.2}
    \begin{pmatrix}
        \lambda_{\ell}\\
        \tfrac{1}{4}(3\lambda_{\ell}+\lambda_h)\\
        \tfrac{1}{2}(\lambda_{\ell}+\lambda_h)\\
        \tfrac{1}{4}(\lambda_{\ell}+3\lambda_h)\\
        \lambda_h
    \end{pmatrix}
    \label{eq:Lambda_e_and_m4-1}
\end{align}
and
\begin{align}
    \vec \Lambda_{\rm e,i} = \frac{1}{\Omega_{\rm e,LIN}}
    \renewcommand{\arraystretch}{1.2}
    \begin{pmatrix}
        \bar p_h^4\lambda_{\ell}\\
        p_h\bar p_h^3(3\lambda_{\ell}+\lambda_h)\\
        3p_h^2\bar p_h^2(\lambda_{\ell}+\lambda_h)\\
        p_h^3\bar p_h(\lambda_{\ell}+3\lambda_h)\\
        p_h^4\lambda_h 
    \end{pmatrix}.
    \label{eq:Lambda_e_and_m4-2}
\end{align}
Using $B_{\rm e}$ and $D_{\rm e}$ given by Eqs.~\eqref{eq:Be_De_m4-1} and \eqref{eq:Be_De_m4-2}, respectively, we obtain 
\begin{align}
    A_{\rm e,LIN}(t_{\rm d})=\frac{(\lambda_h-\lambda_{\ell})^2 p_h\bar p_h\eta^{t_{\rm d}}}{4(p_h\lambda_h+\bar p_h\lambda_{\ell})(1-p_h\lambda_h-\bar p_h\lambda_{\ell})}.
    \label{eq:acf_e_lin_m4}
\end{align}
These results for $A_{\rm e,LIN}(t_{\rm d})$ with $m=2,3,4$ can be summarized into Eq.~\eqref{eq:acf_e_lin_m}.

\section{Derivation of ACF for a node in some simple cases}\label{append:acf_node_simple}

Here we derive the ACF for a node belonging to $k$ hyperedges of sizes $\vec m=(m_1,\ldots,m_k)$. We remind that $v(t)=1$ if at least one of the $k$ hyperedges generates an event at time $t$ and $v(t)=0$ otherwise [see Eq.~\eqref{eq:vt_ejt_relation}]. We reuse the notations introduced in the main text. 
In the same manner as the derivation of Eq.~\eqref{eq:acf_e_define_simple}, we find that the ACF reads
\begin{align}
    A_{\rm v}(t_{\rm d})\equiv \frac{ \langle v(t)v(t+t_{\rm d})\rangle- \langle v(t)\rangle^2}{ \langle v(t)^2\rangle- \langle v(t)\rangle^2}=\frac{ R_{\rm v}(t_{\rm d})- \Omega_{\rm v}}{1-\Omega_{\rm v}},
    \label{eq:acf_v_define}
\end{align}
where
\begin{align}
    R_{\rm v}(t_{\rm d})\equiv \Pr[v(t+t_{\rm d})=1|v(t)=1].
    \label{eq:Rtd_n_define}
\end{align}
To derive $R_{\rm v}(t_{\rm d})$, we set $t=0$ without loss of generality in Eq.~\eqref{eq:Rtd_n_define} to obtain
\begin{align}
    R_{\rm v}(t_{\rm d})=\Pr[v(t_{\rm d})=1|v(0)=1]= 
    \sum_{\vec n,\vec n'} 
    \Pr[v(t_{\rm d})=1|\vec \mu_h(t_{\rm d})=\vec n]
    \Pr[\vec \mu_h(t_{\rm d})=\vec n|\vec \mu_h(0)=\vec n']
    \Pr[\vec \mu_h(0)=\vec n'|v(0)=1],
    \label{eq:Rtd_v_define}
\end{align}
where $\vec n=(n_{0},n_{1},\ldots,n_{k})$ with $0\leq n_{j}\leq \mu_j$ for $j\in\{0,1,\ldots,k\}$. Using the same notations as those in Eq.~\eqref{eq:Ptau_v_matrix_append}, we write
\begin{align}
    R_{\rm v}(t_{\rm d})= \vec \Lambda_{\rm v,f}^\top W_{\rm v}^{t_{\rm d}}\vec \Lambda_{\rm v,i}, 
    \label{eq:Rv_td_matrix}
\end{align}
enabling us to derive $R_{\rm v}(t_{\rm d})$ and hence $A_{\rm v}(t_{\rm d})$ in Eq.~\eqref{eq:acf_v_define}. We note that matrix $W_{\rm v}$ in Eq.~\eqref{eq:Rv_td_matrix} can be diagonalized into
\begin{equation}
    W_{\rm v}=B_{\rm v}D_{\rm v}B_{\rm v}^{-1},
    \label{eq:Wv_BDB1}
\end{equation}
where
\begin{align}
    B_{\rm v}\equiv& \bigotimes_{j=0}^k B_{{\rm e},j},
    \label{eq:BDB1_v-1}\\
    D_{\rm v}\equiv& \bigotimes_{j=0}^k D_{{\rm e},j}
    \label{eq:BDB1_v-2}
\end{align}
and
\begin{align}
    B_{\rm v}^{-1}\equiv& \bigotimes_{j=0}^k B_{{\rm e},j}^{-1}.
    \label{eq:BDB1_v-3}
\end{align}
Here $D_{{\rm e},j}$ and $B_{{\rm e},j}$ for each $j$ are given by Eqs.~\eqref{eq:De_matrix} and~\eqref{eq:Be_element}, respectively.

To demonstrate the analytically derived ACF, we consider a node belonging to $k$ hyperedges of the same size $m$, i.e., $m_1=\ldots=m_k=m$. When $k=2$ and $m_1=m_2=2$, one has $\mu_0=\mu_1=\mu_2=1$ and $\mu_{0h},\mu_{1h},\mu_{2h}\in \{0,1\}$ such that $\vec\mu_h\in \{0,1\}^3$, or equivalently, $\vec\mu_h\in \{(0,0,0),(0,0,1),\ldots,(1,1,0),(1,1,1)\}$. Equations~\eqref{eq:Pr_1n_node} and~\eqref{eq:Pr_n0} in this case read
\begin{equation}
    \Pr[v(t)=1|\vec \mu_h(t)=(0,0,0)] = 1-\bar\lambda_{\ell}^2,\ \ldots,\ \Pr[v(t)=1|\vec \mu_h(t)=(1,1,1)] = 1-\bar\lambda_h^2 
    \label{eq:Pr_1n_node_and}
\end{equation}
and
\begin{equation}
    \Pr[\vec \mu_h(0)=(0,0,0)]=\bar p_h^3,\ldots,\Pr[\vec \mu_h(0)=(1,1,1)]=p_h^3, \label{eq:fmn_node}
\end{equation}
respectively, leading to
\begin{align}
    \vec\Lambda_{\rm v,f} = 
    \renewcommand{\arraystretch}{1.2}
    \begin{pmatrix}
 1 - \bar\lambda_{\ell}^2 \\
 1 - \bar\lambda_{\ell}^2 \\
 1 - \bar\lambda_{\ell}^2 \\
 1 - \bar\lambda_{\ell}^2 \\
 1 - \bar\lambda_{\ell}^2 \\
 1-\bar\lambda_h\bar\lambda_{\ell} \\
 1-\bar\lambda_h\bar\lambda_{\ell} \\
 1 - \bar\lambda_h^2
\end{pmatrix}
\label{eq:Lambda_v_AND_k2m2-1}
\end{align}
and
\begin{align}
\vec\Lambda_{\rm v,i} = \frac{1}{\Omega_{\rm v,AND}}
\renewcommand{\arraystretch}{1.2}
\begin{pmatrix} 
    \bar p_h^3(1-\bar\lambda_{\ell}^2)\\
    p_h\bar p_h^2(1-\bar\lambda_{\ell}^2)\\
    p_h\bar p_h^2(1-\bar\lambda_{\ell}^2)\\
    p_h^2\bar p_h(1-\bar\lambda_{\ell}^2)\\
    p_h\bar p_h^2(1-\bar\lambda_{\ell}^2)\\
    p_h^2\bar p_h(1-\bar\lambda_h\bar\lambda_{\ell})\\
    p_h^2\bar p_h(1-\bar\lambda_h\bar\lambda_{\ell})\\
    p_h^3(1-\bar\lambda_h^2)
    \end{pmatrix}.
    \label{eq:Lambda_v_AND_k2m2-3}
\end{align}
The transition probability sub-matrices are given by
\begin{align}
    W_{{\rm e},0}=W_{{\rm e},1}=W_{{\rm e},2}=\begin{pmatrix}
        \bar r_{\ell h} & r_{h \ell}\\
        r_{\ell h} & \bar r_{h \ell}
    \end{pmatrix},
    \label{eq:Wej_k2m2}
\end{align}
which are diagonalized into $W_{{\rm e},j}=B_{{\rm e},j}D_{{\rm e},j}B_{{\rm e},j}^{-1}$, $j \in \{0,1,2 \}$ with
\begin{equation}
    B_{{\rm e},j}=\begin{pmatrix}
        \alpha & -1\\ 1 & 1
    \end{pmatrix}
\label{eq:B_e-2x2}
\end{equation}
and
\begin{equation}
    D_{{\rm e},j}=\begin{pmatrix}
        1 & 0 \\ 0 & \eta
    \end{pmatrix}.
\label{eq:D_e-2x2}
\end{equation}
By substituting Eqs.~\eqref{eq:B_e-2x2} and \eqref{eq:D_e-2x2} in
Eqs.~\eqref{eq:BDB1_v-1} and \eqref{eq:BDB1_v-2}, respectively, one obtains
\begin{equation}
    B_{\rm v}=\begin{pmatrix}
 \alpha^3 & -\alpha^2 & -\alpha^2 & \alpha & -\alpha^2 & \alpha & \alpha & -1 \\
 \alpha^2 & \alpha^2 & -\alpha & -\alpha & -\alpha & -\alpha & 1 & 1 \\
 \alpha^2 & -\alpha & \alpha^2 & -\alpha & -\alpha & 1 & -\alpha & 1 \\
 \alpha & \alpha & \alpha & \alpha & -1 & -1 & -1 & -1 \\
 \alpha^2 & -\alpha & -\alpha & 1 & \alpha^2 & -\alpha & -\alpha & 1 \\
 \alpha & \alpha & -1 & -1 & \alpha & \alpha & -1 & -1 \\
 \alpha & -1 & \alpha & -1 & \alpha & -1 & \alpha & -1 \\
 1 & 1 & 1 & 1 & 1 & 1 & 1 & 1
\end{pmatrix}
\label{eq:Bv_Dv_k2m2-1}
\end{equation}
and
\begin{equation}
    D_{\rm v}=\begin{pmatrix}
 1 & 0 & 0 & 0 & 0 & 0 & 0 & 0 \\
 0 & \eta & 0 & 0 & 0 & 0 & 0 & 0 \\
 0 & 0 & \eta & 0 & 0 & 0 & 0 & 0 \\
 0 & 0 & 0 & \eta^2 & 0 & 0 & 0 & 0 \\
 0 & 0 & 0 & 0 & \eta & 0 & 0 & 0 \\
 0 & 0 & 0 & 0 & 0 & \eta^2 & 0 & 0 \\
 0 & 0 & 0 & 0 & 0 & 0 & \eta^2 & 0 \\
 0 & 0 & 0 & 0 & 0 & 0 & 0 & \eta^3
\end{pmatrix},
\label{eq:Bv_Dv_k2m2-2}
\end{equation}
respectively. Using $\Omega_{\rm v,AND}$ in Eq.~\eqref{eq:Omega_v_and_k2m} with $m=2$ and Eqs.~\eqref{eq:Lambda_v_AND_k2m2-1},~\eqref{eq:Lambda_v_AND_k2m2-3},~\eqref{eq:Bv_Dv_k2m2-1}, and~\eqref{eq:Bv_Dv_k2m2-2}, we obtain 
\begin{align}
    A_{\rm v,AND}(t_{\rm d})=\frac{(\lambda_h-\lambda_{\ell})^2 p_h\bar p_h}{\Omega_{\rm v,AND}(1-\Omega_{\rm v,AND})}\sum_{n=1}^3 c_n\eta^{nt_{\rm d}},
    \label{eq:acf_v_and_k2m2}
\end{align}
where
\begin{align}
    c_1 &\equiv p_h^2 \left[
    3p_h^2\bar\lambda_h^2
    -2p_h (3p_h-4)\bar\lambda_h\bar\lambda_{\ell}
    + (3p_h^2-8p_h+6)\bar\lambda_{\ell}^2
    \right],\\
    c_2 &\equiv p_h\bar p_h \left[
    3p_h^2\bar\lambda_h^2
    -2p_h (3p_h-2)\bar\lambda_h\bar\lambda_{\ell}
    + (3p_h^2-4p_h+2)\bar\lambda_{\ell}^2
    \right],\\
    c_3 &\equiv p_h^2 \bar p_h^2(\lambda_h-\lambda_{\ell})^2.
\end{align}

We also derive the analytical solution for a node with $k=2$ and $m=3$, i.e., $m_1=m_2=3$, as follows:
\begin{align}
    A_{\rm v,AND}(t_{\rm d})=\frac{(\lambda_h-\lambda_{\ell})^2 p_h\bar p_h}{\Omega_{\rm v,AND}(1-\Omega_{\rm v,AND})}\sum_{n=1}^5 c_n\eta^{nt_{\rm d}},
    \label{eq:acf_v_and_k2m3}
\end{align}
where
\begin{align}
    \Omega_{\rm v,AND}& =1-p_h^5 \bar\lambda_h^2 
    -2p_h^3\bar p_h(p_h+1) \bar\lambda_h\bar\lambda_{\ell}
    -(p_h^4+8p_h^3\bar p_h+10p_h^2\bar p_h^2+5p_h\bar p_h^3+\bar p_h^4)\bar p_h \bar\lambda_{\ell}^2,\\
    c_1 &\equiv p_h^4 \left[
    5p_h^4\bar\lambda_h^2 
    -2p_h^2(5p_h^2-6)\bar\lambda_h \bar\lambda_{\ell}
    +(5p_h^4-12p_h^2+8)\bar\lambda_{\ell}^2
    \right],\\
    c_2 &\equiv p_h^3\bar p_h \left[
    10p_h^4\bar\lambda_h^2 
    -4p_h^2(5p_h^2-3)\bar\lambda_h \bar\lambda_{\ell}
    +2(5p_h^4-6p_h^2+3)\bar\lambda_{\ell}^2
    \right],\\
    c_3 &\equiv p_h^2\bar p_h^2 \left[
    10p_h^4\bar\lambda_h^2 
    -4p_h^2(5p_h^2-1)\bar\lambda_h \bar\lambda_{\ell}
    +2(5p_h^4-2p_h^2+1)\bar\lambda_{\ell}^2
    \right],\\
    c_4 &\equiv 5p_h^5 \bar p_h^3(\lambda_h-\lambda_{\ell})^2,\\
    c_5 &\equiv p_h^4 \bar p_h^4(\lambda_h-\lambda_{\ell})^2.
\end{align}

For a node with $k=3$ and $m=2$, i.e., $m_1=m_2=m_3=2$, we obtain
\begin{align}
    A_{\rm v,AND}(t_{\rm d})=\frac{(\lambda_h-\lambda_{\ell})^2 p_h\bar p_h}{\Omega_{\rm v,AND}(1-\Omega_{\rm v,AND})}\sum_{n=1}^4 c_n\eta^{nt_{\rm d}},
    \label{eq:acf_v_and_k3m2}
\end{align}
where 
\begin{align}
    \Omega_{\rm v,AND}& =1-p_h^4 \bar\lambda_h^3 
    -3p_h^3\bar p_h \bar\lambda_h^2\bar\lambda_{\ell}
    -3p_h^2\bar p_h^2 \bar\lambda_h\bar\lambda_{\ell}^2
    -(p_h^3+3p_h^2\bar p_h+4p_h\bar p_h^2+\bar p_h^3)\bar p_h\bar\lambda_{\ell}^3,\\
    c_1 &\equiv p_h^2 \left[
    4p_h^4\bar\lambda_h^4 
    -2p_h^3(8p_h-9)\bar\lambda_h^3 \bar\lambda_{\ell}
    +3p_h^2(8p_h^2-18p_h+11)\bar\lambda_h^2 \bar\lambda_{\ell}^2
    -2p_h(8p_h^3-27p_h^2+33p_h-15)\bar\lambda_h\bar\lambda_{\ell}^3 \right.\notag\\
    &\left. +(4p_h^4-18p_h^3+33p_h^2-30p_h+12)\bar\lambda_{\ell}^4
    \right],\\
    c_2 &\equiv p_h\bar p_h \left[
    6p_h^4\bar\lambda_h^4 
    -6p_h^3(4p_h-3)\bar\lambda_h^3 \bar\lambda_{\ell}
    +3p_h^2(12p_h^2-18p_h+7)\bar\lambda_h^2 \bar\lambda_{\ell}^2
    -6p_h(4p_h^3-9p_h^2+7p_h-2)\bar\lambda_h \bar\lambda_{\ell}^3 \right.\notag\\
    &\left.+3(2p_h^4-6p_h^3+7p_h^2-4p_h+1)\bar\lambda_{\ell}^4
    \right],\\
    c_3 &\equiv p_h^2\bar p_h^2 \left[
    4p_h^2\bar\lambda_h^4
    -2p_h(8p_h-3)\bar\lambda_h^3 \bar\lambda_{\ell}
    +3(8p_h^2-6p_h+1)\bar\lambda_h^2 \bar\lambda_{\ell}^2
    -2(8p_h^2-9p_h+3)\bar\lambda_h \bar\lambda_{\ell}^3 \right.\notag\\
    &\left.+(4p_h^2-6p_h+3)\bar\lambda_{\ell}^4
    \right],\\
    c_4 &\equiv p_h^3 \bar p_h^3(\lambda_h-\lambda_{\ell})^4.
\end{align}

Next, to derive the ACF under the LIN rule, we again consider a focal node belonging to two hyperedges of size two (i.e., two edges), i.e., $k=2$ and $m_1=m_2=m=2$. We obtain
\begin{align}
    \vec\Lambda_{\rm v,f} =
    \renewcommand{\arraystretch}{1.2}
    \begin{pmatrix} 
    1-\bar\lambda_{\ell}^2\\
    1-\tfrac{1}{2}\bar\lambda_{\ell}(\bar\lambda_h+\bar\lambda_{\ell})\\
    1-\tfrac{1}{2}\bar\lambda_{\ell}(\bar\lambda_h+\bar\lambda_{\ell})\\
    1-\tfrac{1}{4}(\bar\lambda_h+\bar\lambda_{\ell})^2\\
    1-\tfrac{1}{4}(\bar\lambda_h+\bar\lambda_{\ell})^2\\
    1-\tfrac{1}{2}(\bar\lambda_h+\bar\lambda_{\ell})\bar\lambda_h\\
    1-\tfrac{1}{2}(\bar\lambda_h+\bar\lambda_{\ell})\bar\lambda_h\\
    1-\bar\lambda_h^2
    \end{pmatrix}
\label{eq:Lambda_v_LIN_k2m2-1}
\end{align}
and
\begin{align}
    \vec\Lambda_{\rm v,i} =
    \frac{1}{\Omega_{\rm v,LIN}}
    \renewcommand{\arraystretch}{1.2}
    \begin{pmatrix} 
    \bar p_h^3(1-\bar\lambda_{\ell}^2)\\
    p_h\bar p_h^2\left[1-\tfrac{1}{2}\bar\lambda_{\ell}(\bar\lambda_h+\bar\lambda_{\ell})\right]\\
    p_h\bar p_h^2\left[1-\tfrac{1}{2}\bar\lambda_{\ell}(\bar\lambda_h+\bar\lambda_{\ell})\right]\\
    p_h^2\bar p_h\left[1-\tfrac{1}{4}(\bar\lambda_h+\bar\lambda_{\ell})^2\right]\\
    p_h\bar p_h^2\left[1-\tfrac{1}{4}(\bar\lambda_h+\bar\lambda_{\ell})^2\right]\\
    p_h^2\bar p_h\left[1-\tfrac{1}{2}(\bar\lambda_h+\bar\lambda_{\ell})\bar\lambda_h\right]\\
    p_h^2\bar p_h\left[1-\tfrac{1}{2}(\bar\lambda_h+\bar\lambda_{\ell})\bar\lambda_h\right]\\
    p_h^3(1-\bar\lambda_h^2)
    \end{pmatrix}.
    \label{eq:Lambda_v_LIN_k2m2-2}
\end{align}
Using $\Omega_{\rm v,LIN}$ in Eq.~\eqref{eq:Omega_v_lin_k2m} with $m=2$ and Eqs.~\eqref{eq:Bv_Dv_k2m2-1},~\eqref{eq:Bv_Dv_k2m2-2},~\eqref{eq:Lambda_v_LIN_k2m2-1}, and~\eqref{eq:Lambda_v_LIN_k2m2-2}, we obtain 
\begin{align}
    A_{\rm v,LIN}(t_{\rm d})=\frac{(\lambda_h-\lambda_{\ell})^2 p_h\bar p_h}{\Omega_{\rm v,LIN}(1-\Omega_{\rm v,LIN})}\sum_{n=1}^2 c'_n\eta^{nt_{\rm d}},
    \label{eq:acf_v_lin_k2m2}
\end{align}
where
\begin{align}
    c'_1 &\equiv \tfrac{1}{16} \left[
    (12p_h^2+4p_h+1)\bar\lambda_h^2
    -6(4p_h^2-4p_h-1)\bar\lambda_h\bar\lambda_{\ell}
    +(12p_h^2-28p_h+17)\bar\lambda_{\ell}^2 \right],\\
    c'_2 &\equiv \tfrac{3}{16} p_h \bar p_h(\lambda_h-\lambda_{\ell})^2.
\end{align}

We also derive the analytical solution for a node with $k=2$ and $m=3$, i.e., $m_1=m_2=3$, as follows:
\begin{align}
    A_{\rm v,LIN}(t_{\rm d})=\frac{(\lambda_h-\lambda_{\ell})^2 p_h\bar p_h}{\Omega_{\rm v,LIN}(1-\Omega_{\rm v,LIN})}\sum_{n=1}^2 c'_n\eta^{nt_{\rm d}},
    \label{eq:acf_v_lin_k2m3}
\end{align}
where
\begin{align}
\Omega_{\rm v,LIN} &=1 -\tfrac{1}{9}p_h(8p_h+1)\bar\lambda_h^2
    -\tfrac{16}{9}p_h\bar p_h \bar\lambda_h\bar\lambda_{\ell}
    -\tfrac{1}{9}\bar p_h(8\bar p_h+1)\bar\lambda_{\ell}^2,\\
    c'_1 &\equiv \tfrac{1}{81}\left[
    (52p_h^2+8p_h+1)\bar\lambda_h^2
    -2(52p_h^2-52p_h-5)\bar\lambda_h\bar\lambda_{\ell}
    +(52p_h^2-112p_h+61)\bar\lambda_{\ell}^2\right],\\
    c'_2 &\equiv \tfrac{8}{81} p_h \bar p_h(\lambda_h-\lambda_{\ell})^2.
\end{align}

For a node with $k=3$ and $m=2$, i.e., $m_1=m_2=m_3=2$, we obtain
\begin{align}
    A_{\rm v,LIN}(t_{\rm d})=\frac{(\lambda_h-\lambda_{\ell})^2 p_h\bar p_h}{\Omega_{\rm v,LIN}(1-\Omega_{\rm v,LIN})}\sum_{n=1}^3 c'_n\eta^{nt_{\rm d}},
    \label{eq:acf_v_lin_k3m2}
\end{align}
where
\begin{align}
    \Omega_{\rm v,LIN} &=1 -p_h(4p_h^2+3p_h+1)\bar\lambda_h^3 -3p_h\bar p_h(4p_h+1)\bar\lambda_h^2\bar\lambda_{\ell} -3p_h\bar p_h(4\bar p_h+1)\bar\lambda_h\bar\lambda_{\ell}^2 -\bar p_h(4\bar p_h^2+3\bar p_h+1)\bar\lambda_{\ell}^3,\\
    c'_1 &\equiv \tfrac{1}{64}\left[
    (36p_h^4+36p_h^3+18p_h^2+6p_h+1)\bar\lambda_h^4 
    -4(36p_h^4-18p_h^3-21p_h^2-9p_h-2) \bar\lambda_h^3\bar\lambda_{\ell}\right.\notag\\
    &+6(36p_h^4-72p_h^3+20p_h^2+16p_h+5) \bar\lambda_h^2\bar\lambda_{\ell}^2
    -4(36p_h^4-126p_h^3+141p_h^2-39p_h-14) \bar\lambda_h\bar\lambda_{\ell}^3\notag\\
    &\left. +(36p_h^4-180p_h^3+342p_h^2-294p_h+97) \bar\lambda_{\ell}^4
    \right],\\
    c'_2 &\equiv \tfrac{3}{64}p_h \bar p_h \left[
    (8p_h^2+4p_h+1)\bar\lambda_h^4 
    -4(8p_h^2-2p_h-1) \bar\lambda_h^3\bar\lambda_{\ell} +2(24p_h^2-24p_h+1) \bar\lambda_h^2\bar\lambda_{\ell}^2
    \right.\notag\\
    &\left. -4(8p_h^2-14p_h+5) \bar\lambda_h\bar\lambda_{\ell}^3 +(8p_h^2-20p_h+13) \bar\lambda_{\ell}^4
    \right],\\
    c'_3 &\equiv \tfrac{1}{16} p_h^2 \bar p_h^2(\lambda_h-\lambda_{\ell})^4.
\end{align}

\section{Numbers of hyperedges analyzed for Figure 6}\label{append:fig6}

Here we provide information on the numbers of hyperedges analyzed for the calculation of $\Omega_{\rm e}$, CV$_{\rm e}$, and $A_{\rm e}(t_{\rm d})$ in Fig.~6 of the main text. Table~\ref{suppletab:data} shows those numbers for $\Omega_{\rm e}$ and CV$_{\rm e}$. The numbers of hyperedges used for computing $A_{\rm e}(t_{\rm d})$ are not necessarily the same  for different values of $t_{\rm d}$. Therefore we provide them in detail in Supplementary Data.

\begin{table}[!h]
\centering
\caption{Numbers of hyperedges analyzed for the calculation of $\Omega_{\rm e}$ and CV$_{\rm e}$ in Fig.~6 of the main text.}
\begin{tabular}{r|r|r|r|r|r|r}
\hline
$m$ & High~school & Primary~school & DBLP & DAWN & NDC-class & NDC-substance\\
\hline
2 & 2270 & 3118 & 11113 & 5215 & 169 & 222 \\ 
3 & 299 & 338 & 2513 & 2705 & 59 & 84 \\ 
4 & 24 & 7 & 411 & 841 & 52 & 51 \\ 
5 & & & 57 & 230 & 39 & 20 \\ 
6 & & & 9 & 33 & 34 & 11 \\ 
7 & & & 2 & 1 & 17 & 1 \\ 
8 & & & 1 & & 9 & 2 \\ 
9 & & & & & 7 &  \\ 
10 & & & & & 4 & 2 \\ 
11 & & & & & 1 & \\ 
12 & & & & & 2 & 1 \\ 
13 & & & & & 2 & 1 \\ 
14 & & & & & 4 & 2 \\ 
15 & & & & & 2 & \\ 
16 & & & & & & 2 \\ 
20 & & & & & 1 & 2 \\ 
21 & & & & & & 1 \\ 
23 & & & & & & 2 \\ 
25 & & & & & & 1 \\ 
\hline
\end{tabular}
\label{suppletab:data}
\end{table}

\section{Effects of $m$-dependent $\lambda_h$ and $\lambda_{\ell}$ on hyperedge activities}\label{append:m_depend}

\begin{figure}[!h]
    \centering
    \includegraphics[width=0.5\textwidth]{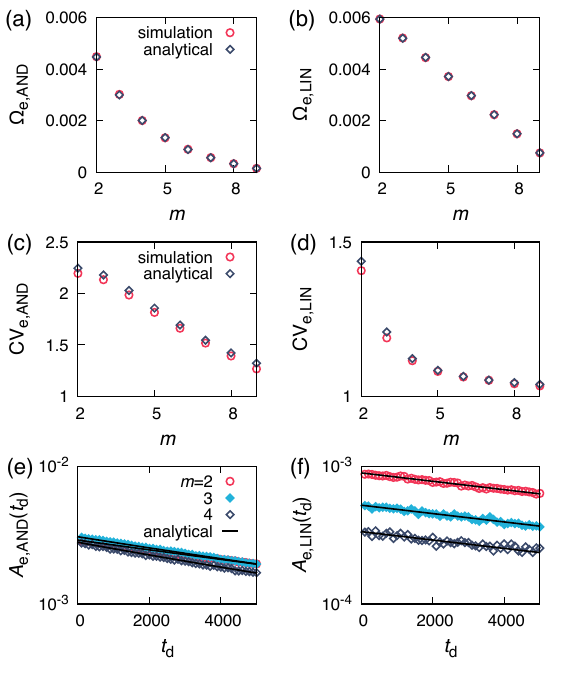}
    \caption{Event-time statistics for hyperedges under $m$-dependent $\lambda_h(m)$ and $\lambda_{\ell}(m)$ given by Eqs.~\eqref{eq:lambda_h_m} and~\eqref{eq:lambda_l_m}: average event probabilities ((a) and (b)), CVs of IETs ((c) and (d)), and ACFs ((e) and (f)) under AND ((a), (c), and (e)) and LIN ((b), (d), and (f)) rules. We generate up to $10^3$ event sequences for the numerical results (shown by circles in panels (a)--(d) and symbols in panels (e) and (f)) with $r_{\ell h}=5\times 10^{-5}$, $r_{h \ell}=2\times 10^{-5}$, $\lambda_{h0}=10^{-2}$, $c_h=-10^{-3}$, $\lambda_{\ell 0}=10^{-3}$, and $c_{\ell}=-10^{-4}$, following the numerical procedures described in the Methods section in the main text. For the analytical results (shown by diamonds in panels (a)--(d) and solid lines in panels (e) and (f)), we numerically calculate Eqs.~\eqref{eq:Omega_e_and_mm} and~\eqref{eq:Omega_e_lin_mm}, as well as Eqs.~\eqref{eq:cv_e_and},~\eqref{eq:cv_e_lin},~\eqref{eq:acf_e_and_m}, and~\eqref{eq:acf_e_lin_m} after replacing $\lambda_h$ and $\lambda_{\ell}$ by $\lambda_h(m)$ and $\lambda_{\ell}(m)$ in Eqs.~\eqref{eq:lambda_h_m} and~\eqref{eq:lambda_l_m}, respectively.
    }
    \label{fig:m-depend}
\end{figure}

\begin{figure}[!h]
    \centering
    \includegraphics[width=0.6\textwidth]{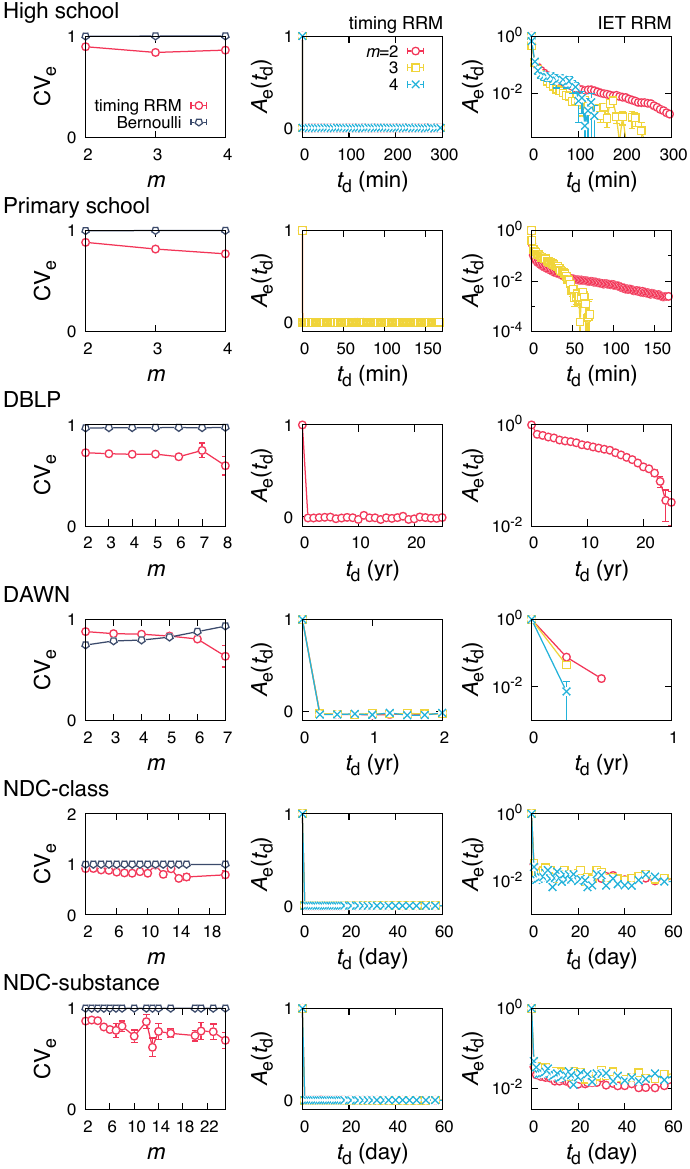}
    \caption{Results for two randomized reference models (RRMs) of event sequences on hyperedges: the timing RRM  and the IET RRM. We generate 10 randomized event sequences for each hyperedge included in Fig.~6 of the main text to calculate their averaged CVs and ACFs using the timing RRM (left and central columns) and ACFs using the IET RRM (right columns). Error bars stand for standard errors.
    }
    \label{fig:null}
\end{figure}

So far, and in the main text, we have assumed that event probabilities $\lambda_h$ and $\lambda_{\ell}$ are independent of the size of the hyperedge, $m$. Here we consider $m$-dependent $\lambda_h$ and $\lambda_{\ell}$ by assuming a linear dependence as follows:
\begin{align}
    \lambda_h(m)&=\lambda_{h0}+c_h m, \label{eq:lambda_h_m}\\
    \lambda_{\ell}(m)&=\lambda_{\ell 0}+c_{\ell} m,
    \label{eq:lambda_l_m}
\end{align}
where $\lambda_{h0}$, $c_h$, $\lambda_{\ell 0}$, and $c_{\ell}$ are constants. For the AND rule, we obtain the event probability on the hyperedge as
\begin{align}
    \lambda_{\rm e}(m,m_h) = \delta_{m,m_h}(\lambda_{h0} + c_h m) +(1-\delta_{m,m_h})(\lambda_{\ell 0}+c_{\ell}m),
    \label{eq:event_rate_and_m}
\end{align}
where $m_h \in \{0,\ldots,m\}$ is the number of nodes in the $h$ state in the hyperedge. Under the LIN rule, the event probability on the hyperedge is given by
\begin{align}
    \lambda_{\rm e}(m,m_h)=\frac{m_h}{m}(\lambda_{h0} + c_h m)+\left(1-\frac{m_h}{m}\right)(\lambda_{\ell 0}+c_{\ell}m).
    \label{eq:event_rate_lin_m}
\end{align}

We substitute Eqs.~\eqref{eq:event_rate_and_m} and~\eqref{eq:event_rate_lin_m} into the analytical results for the average event probability on the hyperedge in Eq.~\eqref{eq:Omega_e} to obtain
\begin{align}
    \Omega_{\rm e,AND}=p_h^m \lambda_{h0} +(1-p_h^m)\lambda_{\ell 0} + m [p_h^m c_h +(1-p_h^m)c_{\ell}]
    \label{eq:Omega_e_and_mm}
\end{align}
and
\begin{align}
    \Omega_{\rm e,LIN}=p_h\lambda_{h0}+\bar p_h\lambda_{\ell 0}+m[p_hc_{h}+\bar p_hc_{\ell}].
    \label{eq:Omega_e_lin_mm}
\end{align}
We observe that $\Omega_{\rm e,LIN}$ is now a linear function of $m$, which was not the case when $\lambda_h$ and $\lambda_{\ell}$ were constant. Considering the empirical results that the average event probability overall decreases with $m$, one can choose parameter values satisfying that $p_h c_h+\bar p_h c_{\ell}<0$; we consider the case with $c_h,c_{\ell}<0$. Specifically, for both AND and LIN rules, we set $\lambda_{h0}=10^{-2}$, $c_h=-10^{-3}$, $\lambda_{\ell 0}=10^{-3}$, and $c_{\ell}=-10^{-4}$, for the range of $2\leq m\leq 9$, guaranteeing the condition that $\lambda_h(m)>\lambda_{\ell}(m)>0$. As shown in Fig.~\ref{fig:m-depend}(a,b), both $\Omega_{\rm e,AND}$ and $\Omega_{\rm e,LIN}$ decrease with $m$, implying that the LIN rule cannot be excluded as a mechanism behind the empirical observations.

In addition to the average event probability, the CV of IETs in Eq.~\eqref{eq:cv_e_m} and the ACF in Eqs.~\eqref{eq:acf_e_and_m} and~\eqref{eq:acf_e_lin_m} can be analytically derived and numerically estimated using $\lambda_h(m)$ and $\lambda_{\ell}(m)$ in Eqs.~\eqref{eq:lambda_h_m} and~\eqref{eq:lambda_l_m}. The results are shown in Fig.~\ref{fig:m-depend}(c--f).

\section{Randomized reference models for data analysis}\label{append:null}

For the empirical data analysis, we consider two randomized reference models (RRMs)~\cite{Gauvin2022Randomized, Jo2020Bursttree}, namely, timing randomization and IET shuffling. The main idea is that we randomize some features of temporal patterns on hyperedges, while retaining all other features to see if the randomized features can account for the temporal patterns. Let us consider an empirical event sequence on the hyperedge with $n$ events in the time range $\{1,\ldots,T\}$.

For the timing randomization (timing RRM), we assign to each event a random timing $t\in\{1,\ldots,T\}$, implying a Bernoulli process. We preserve the number of events on $\{1,\ldots,T\}$ upon randomization. Then, we expect that 
\begin{align}
{\rm CV}_{\rm e}&=\sqrt{1-\frac{n}{T}},\label{eq:timing-RRM-CV}\\
A_{\rm e}(t_{\rm d})&=\delta_{t_{\rm d},0},\label{eq:timing-RRM-ACF}
\end{align}
irrespective of $m$. For each dataset, we generate 10 randomized event sequences for each hyperedge to calculate their CVs and ACFs. As shown in the left (CV) and central (ACF) columns of Fig.~\ref{fig:null}, the behavior of CVs and ACFs is consistent with Eqs.~\eqref{eq:timing-RRM-CV} and \eqref{eq:timing-RRM-ACF}, respectively, including the independence of $m$.

For the IET shuffling (IET RRM), we permute the IETs in a uniformly random order to construct the event sequence using the shuffled IET sequence. By doing so, we destroy correlations between IETs if any. Since IETs are retained by the IET shuffling, the CV of the shuffled IET sequence remains the same as that of the original data. Therefore, we do not investigate the CV for the IET RRM in the following data analysis. For each dataset, we generate 10 randomized event sequences for each hyperedge to calculate their ACFs. We find in the right columns of Fig.~\ref{fig:null} that most results show overall similar patterns to the original ones, implying that the correlations between IETs appear to be negligible in the original empirical data.

%